\documentclass[aps,prd,reprint,groupedaddress,amsmath]{revtex4-2}

\usepackage[colorlinks,linkcolor=blue]{hyperref}
\usepackage{graphicx}
\usepackage{dcolumn}
\usepackage{bm}
\usepackage{footnote}

\begin{document}


\title{The Gravitational Deflection of Acoustic Charged Black Hole}


\author{Chen-Kai Qiao}
\email{chenkaiqiao@cqut.edu.cn}
\affiliation{College of Science, Chongqing University of Technology, Banan, Chongqing, 400054, China}

\author{Mi Zhou}
\email{lilyzm@cqut.edu.cn}
\affiliation{College of Science, Chongqing University of Technology, Banan, Chongqing, 400054, China}


\date{\today}

\begin{abstract}
Acoustic black hole is becoming an attractive topic in recent years, for it open-up new direction for experimental / observational explorations of black holes. In this work, the gravitational bending of acoustic Schwarzschild black hole is investigated. The gravitational deflection angle of particles traveling along null geodesics, weak gravitational lensing and Einstein ring for acoustic Schwarzschild black hole are carefully studied and analyzed. Particularly, in the calculation of gravitational deflection angle, we resort to two approaches --- the Gauss-Bonnet theorem and the geodesic method. The results show that the gravitational bending effect in acoustic Schwarzschild black hole is enhanced, compared with conventional Schwarzschild black hole. This result indicates that the acoustic black holes may be more easily detectable in gravitational bending effects and weak gravitational lensing observations.
\end{abstract}


\maketitle

\section{Introduction \label{sec:1}}

Gravitational bending is one of the most important phenomenon predicted by Einstein's general theory of relativity. This prediction was first observed in 1919 by A. S. Eddington \emph{et al.} as a demonstration of general relativity \cite{Dyson1919}. As the significant application of gravitational bending, the weak gravitational lensing is becoming an important tool in physics, astronomy and cosmology \cite{Wambsganss1998,Bartelmann2001}. The gravity theories and astrophysical models can be tested from the gravitational bending effects and weak gravitational lensing observations \cite{Wambsganss1998,Bartelmann2001,Uitert2012,Virbhadra2000}. The dark matter distribution in galaxies can also be revealed from gravitational lensing \cite{Uitert2012}. Furthermore, the circular motion of particles and black hole shadows are closely connected to the gravitational lensing \cite{Cunha2018,Mustafa2022a,Mustafa2022b,Atamurotov2022a,Atamurotov2022b,GuoGZ2022,Perlick2022}. In the past decades, the gravitational bending effect of spacetime and the gravitational lensing phenomenon become central issues in physics and astronomy.   

Black hole, since it was proposed decades ago, has attracted great interests in high-energy physics, astrophysics and astronomy. Significantly important information on gravitation, thermodynamics and quantum effects in curved spacetime can be revealed from black holes \cite{Black-Hole-Physics}. For a long time, researches on black holes were mainly motivated by theories. However, huge progresses on experimental / observational explorations of black holes have been witnessed in recent years. The gravitational wave signals were detected by LIGO and Virgo from the merging of binary black holes \cite{LIGO2016}. The first black hole image at the center of galaxy M87 was observed by Event Horizon Telescope in 2019 \cite{Akiyama2019}.

Besides the astrophysical side, many investigations on black holes are also taken place in other branches of physics. The emergence of acoustic black hole is one of such attempts. Historically, acoustic black hole was first proposed by W. G. Unruh to provide a connection between astrophysical black holes and tabletop experiments \cite{Unruh1981}. These kind of black holes can be generated by moving fluid with speed faster than the local sound speed. The behavior of fluid eventually generate an analogue gravity model and form an effective curved spacetime. These acoustic black holes and corresponding analogue gravity models become extremely useful in quantum physics, high-energy physics and condensed matter physics \cite{Visser2005}. The interplay between black holes and quantum particles, as well as their quantum gravitational effects, could be mimicked by such kind of black holes \cite{Visser1998,Visser1999,Barcelo2018}. There are a number of progresses in this area in the past years. Experimentally, the acoustic black hole was first reported in Bose-Einstein condensate system \cite{Lahav2010}. And similar experimental realization of acoustic black holes also emerged in other systems \cite{Steinhauer2014,Drori2019,Blencowe2020,Drinkwater2020}. Remarkably, the analogue Hawking radiation and its Hawking temperature of acoustic black holes were also successfully realized in experiments \cite{Drori2019,Nova2019,Isoard2020}. With this superiority, acoustic black holes and corresponding analog gravity models have great impacts on theoretical as well as experimental physics. 

Furthermore, apart from the aforementioned mechanisms, acoustic black holes can also be generated from astrophysical black hole systems with fluids surrounded. Firstly, black holes (such as the Schwarzschild black hole) in the bath of cosmological microwave background is natural candidates for astrophysical acoustic black hole \cite{Ge2019}. Secondly, other studies suggested that relativistic and transonic accretions onto astrophysical black holes would provide examples of astrophysical acoustic black holes \cite{Das2004,Das2006}. In addition, black holes may be surrounded by superfluid dark matter \cite{Berezhiani2015}. The transonic accretion of such superfluid dark matter on black hole systems also provide a scenario realizing analogue black holes \cite{Ge2019}.

The acoustic black holes have extremely rich properties in nature. Because of the equivalent mathematical descriptions between other black holes and acoustic black holes, it can be conjectured that, if a physical phenomenon happens in other black hole scenarios, it may also occur in acoustic black holes and analogue gravity models. Recently, the acoustic black holes have attracted large number of interests. Many kinds of acoustic black holes have been constructed by fluid systems propagating in flat or curved spacetime background \cite{Visser2005,Visser1998,Ge2019,Ge2010,Anacleto2010,Yu2019}. These acoustic black holes not only can be produced in condensed matter systems, but also could be generated by mechanisms in high-energy physics and cosmology. In 2019, X.-H. Ge \emph{et al.} obtained a class of acoustic black hole solutions for analogue gravity models by considering the relativistic Gross-Pitaevskii and Yang-Mills theories \cite{Ge2019}. Furthermore, the horizons, shadows, quasinormal moles, quasibound states, Hawking radiations, Unruh effect, quasiparticle propagation for acoustic black holes are studied and analyzed systematically \cite{Fischer2002,Fischer2003,Cardoso2004,Benone2015,Lima2019,Eskin2019,Guo2020,Vieira2021,Vieira2021b,Ling2021}.

In this paper, we study the gravitational bending effect in spacetime generated by acoustic black holes. A simple and representative example, the acoustic Schwarzschild black hole, is chosen in this work. Similar to the conventional Schwarzschild black hole in general relativity, the acoustic Schwarzschild black hole could also reveal some universal properties of the more complex acoustic black holes. A comprehensive understanding of such black hole could provide useful information and profound insights for other acoustic black holes. 

In the present work, the gravitational deflection angle of particles traveling along null geodesics, weak gravitational lensing, Einstein rings of acoustic Schwarzschild black hole are carefully studied and analyzed in details. Hopefully, the conclusions obtained from acoustic Schwarzschild black hole could give hints for more complex acoustic black holes and various analogue gravity models. In many cases (especially the analogue models in condensed matter systems), particles traveling along null geodesics of acoustic black holes maybe quasi-particles \cite{Visser2005,Visser1998}, such as phonons (sound-waves) and vertex in fluid systems, rather than photons (light beams). This would make the gravitational bending and gravitational lensing in acoustic black holes a little bit different from those in other black holes. However, because of the totally equivalent mathematical descriptions, the approaches dueling with gravitational bending of other black holes are still available in the study of acoustic black holes. 

In the calculation of gravitational deflection angle of particles traveling along null geodesics, we resort to two approaches. One is the Gauss-Bonnet theorem from geometrical topology, in which the gravitational deflection angle is determined by geometric and topological properties in curved space. The other approach is the traditional geodesic method, in which the gravitational deflection angle is obtained by solving the trajectories of null geodesics. The Gauss-Bonnet approach was first developed by G. W. Gibbons and M. C. Werner in 2008 \cite{Gibbons2008}. In their approach, the gravitational deflection angle is calculated by applying the Gauss-Bonnet theorem in optical geometry / optical manifold of black hole spacetimes. From this approach, the gravitational deflection angle is directly connected with the Gauss curvature in optical geometry. Therefore, the Gauss-Bonnet theorem can give us new insights on gravitational bending effects. Recently, Gibbons and Werner's approach has been applied to many gravitational systems, and consistent results with the traditional method have been obtained \cite{Werner2012,Ishihara2016a,Ishihara2016b,Ovgun2018,Jusufi2018,Crisnejo2018a,Jusufi2018epjc,Javed2019epjc,Takizawa2020,Pantig2020,Liu2021,Javed2020,Zhang2021}. Furthermore, some studies suggested that, apart from massless photons, the gravitational bending effects for massive particles near black holes and wormholes can also be correctly analyzed using Gauss-Bonnet theorem \cite{Crisnejo2018b,Crisnejo2019,Li2020a,Li2020b,Jusufi2019epjc}. On the other hand, the geodesic method has a very long history, many famous problems in gravitational bending and gravitational lensing can be solved using this approach \cite{Virbhadra2000,Weinberg1972,Virbhadra1998b,Eiroa2004,Keeton2005,Iyer2007,Virbhadra2009,Kim2021,Jusufi2018c}. After developed for many years, this approach has been tested by large numbers of observations, and it has become a widely adopted approach in physics and astronomy. The geodesic method could duel with both weak and strong gravitational lensings with extremely high precision.

This paper is organized as follows: Section \ref{sec:1} gives an introduction of this work. Section \ref{sec:3} briefly describes the  acoustic Schwarzschild black hole. Section \ref{sec:4} is devoted to the gravitational bending effect of acoustic Schwarzschild black hole. Results and discussions on gravitational deflection angle, weak gravitational lensing and Einstein ring are presented in this section. Summary and conclusions are given in section \ref{sec:5}. In this work, the natural unit $G=c=1$ is adopted.

\section{Acoustic Schwarzschild Black Hole \label{sec:3}}

Acoustic black holes can be proposed in several ways \cite{Unruh1981,Visser1998,Ge2019,Ge2010,Anacleto2010,Yu2019}, not only from laboratory tabletop experiments in condensed matter systems, but also from mechanisms in high-energy physics, astronomy and cosmology. In a recent work, X.-H. Ge \emph{et al.} obtained a class of acoustic black hole solutions using the relativistic Gross-Pitaevskii and Yang-Mills theories \cite{Ge2019}. 

In this section, we choose the acoustic black holes generated in Gross-Pitaevskii theory to shown Ge's formulation. The Gross-Pitaevskii theory can describe the vortex motion in a fluid system in flat and curved spacetimes \cite{Gross1961}. This theory is often used in the studies of Bose-Einstein condensation \cite{Berges2012}, boson star \cite{Chavanis2012}, scalar field dark matter \cite{Bernal2006,Boehmer2007,Rindler-Daller2012} and cosmological evaluation \cite{Rindler-Daller2012,Fukuyama2008}. In Gross-Pitaevskii theory, the action is given by \cite{Ge2019,Gross1961}
\begin{equation}
	S = \int d^{4}x \sqrt{-g}
	\bigg(
	|\partial_{\mu}\varphi|^{2}+m^{2}|\varphi|^{2}-\frac{b}{2}|\varphi|^{4} 
	\bigg)
\end{equation}
Here, $m^{2} \sim (T-T_{C})$ is a parameter depending on the temperature of the fluid system, $b$ is a constant, and $\varphi$ is a complex scalar field expressed as  $\varphi=\sqrt{\rho(\vec{x},t)}e^{i\hat{\theta}(\vec{x},t)}$.  The $T_{c}$ is a critical temperature of the Gross-Pitaevskii theory describing phase transitions. When $T>T_{c}$ the phenomenological parameter $m^{2}$ is positive, while for $T<T_{c}$ it becomes negative. For acoustic black hole solutions, $T$ is actually the Hawking-Unruh temperature \cite{Vieira2021}. The complex scalar field $\varphi$ corresponds to the order parameter in phase transition of the fluid system, and it propagates in a static background spacetime 
\begin{equation}
	ds^{2}_{\text{background}} = g_{tt}dt^{2} +g_{rr}dr^{2} +g_{\theta\theta}d\theta^{2} + g_{\phi\phi}d\phi^{2}
\end{equation}
Here, $g_{tt}$, $g_{rr}$, $g_{\theta\theta}$ and $g_{\phi\phi}$ is the metric of background spacetime. After some calculations and rearrangements of parameters, the propagation of the phase fluctuation $\hat{\theta}$ in complex scalar field $\varphi$ can be expressed using the wave propagation equation in curved spacetime \cite{Ge2019}
\begin{equation}
	\frac{1}{\sqrt{-G}}\partial_{\mu}(\sqrt{-G}G^{\mu\nu}\partial_{\nu}\hat{\theta})=0
\end{equation} 
In this equation, $G_{\mu\nu}$ is the effective spacetime metric tensor describing the phase fluctuation propagation in fluid systems \cite{Ge2019,Guo2020,Vieira2021b,Ling2021}.
\begin{eqnarray}
	ds_{\text{acoustic}}^{2} & = & G_{\mu\nu}dx^{\mu}dx^{\nu} \nonumber \\
	& = & c_{s}\sqrt{c_{s}^{2}-v_{\mu}v^{\mu}} \cdot
	\bigg[
	\frac{c_{s}^{2}-v_{r}v^{r}}{c_{s}^{2}-v_{\mu}v^{\mu}} g_{tt}dt^{2} \nonumber
	\\
	&   &   + \frac{c_{s}^{2}}{c_{s}^{2}-v_{r}v^{r}} g_{rr}dr^{2} 
	+ g_{\theta\theta}d\theta^{2} 
	+ g_{\phi\phi}d\phi^{2}
	\bigg]
\end{eqnarray}
The $v_{\mu}=\partial_{\mu}\hat{\theta}$ can be viewed as the velocity in fluid systems, and $c_{s}$ is a constant defined in references \cite{Ge2019,Guo2020} such that $v_{\mu}v^{\mu}=-2c_{s}^{2}$. From the above equations, it is interesting to see that the behavior of phase fluctuation propagation in fluid system is peculiar, as if it was lived in a curved spacetime. Therefore, we can effectively give a mathematical description of black hole spacetime to this fluid system. In this way, the spacetime metric of acoustic black hole is generated in the Gross-Pitaevskii theory.

In this work, we are interested in a typical and simple acoustic black hole, the acoustic Schwarzschild black hole. From Ge's formulation, it can be generated by fluid propagating in a Schwarzschild background. In this case, the background spacetime metric is the Schwarzschild metric
\begin{eqnarray}
	ds^{2}_{\text{background}} & = & g_{tt}dt^{2} +g_{rr}dr^{2} +g_{\theta\theta}d\theta^{2} + g_{\phi\phi}d\phi^{2} \nonumber
	\\
	& = & - \bigg( 1-\frac{2M}{r} \bigg) dt^{2} + \bigg( 1-\frac{2M}{r} \bigg)^{-1} dr^{2}  \nonumber
	\\
	&   & + r^{2} ( d\theta^{2} + \sin^{2}\theta d\phi^{2} )
\end{eqnarray}
For the velocity of moving fluid, one can introduce a tuning parameter $\xi$ such that the radial component of fluid velocity becomes \cite{Ge2019,Guo2020}
\begin{equation}
	v_{r} \propto \sqrt{\frac{2M\xi}{r}}; \ \ \ \ 
	v^{r}=g^{rr}v_{r} \propto \bigg(1-\frac{2M}{r}\bigg)\sqrt{\frac{2M\xi}{r}}
\end{equation} 
For the fluid velocity $v_{r}$ to be real, the tuning parameter must satisfies $\xi \ge 0$. Based on these relations as well as the normalization $v_{\mu}v^{\mu}=-2c_{s}^{2}$ \cite{Ge2019}, the spacetime metric for acoustic Schwarzschild black hole can eventually be derived as \cite{Ge2019,Guo2020,Vieira2021}
\begin{eqnarray}
	ds_{\text{acoustic}}^{2} & = & G_{\mu\nu}dx^{\mu}dx^{\nu} \nonumber \\
	& = & -f(r)dt^{2}
	+\frac{1}{f(r)}dr^{2}
	+r^{2}
	(d\theta^{2}+\sin^{2}\theta d\phi^{2}) \label{spacetime metric}
\end{eqnarray}
where function $f(r)$ is defined as:
\begin{equation}
	f(r)=\bigg( 1-\frac{2M}{r} \bigg) \cdot
	\bigg[ 1-\xi\frac{2M}{r} \bigg( 1-\frac{2M}{r} \bigg) \bigg] \label{spacetime metric2}
\end{equation}
When $\xi=0$, the above spacetime reduces to the conventional Schwarzschild black hole. When $\xi \to +\infty$, the whole spacetime ($0 \le r \le +\infty$) is inside the acoustic black hole \cite{Guo2020}. Similar to the conventional Schwarzschild black hole in general relativity, the acoustic Schwarzschild black hole may also reflect some universal properties of more complex acoustic black holes. A comprehensive understanding of such black hole could provide useful information and insights for many acoustic black holes.

The horizon of acoustic Schwarzschild black hole is determined via equation
\begin{equation}
	f(r) = \bigg( 1-\frac{2M}{r} \bigg) \cdot
	\bigg[ 1-\xi\frac{2M}{r} \bigg( 1-\frac{2M}{r} \bigg) \bigg] =0
\end{equation}
Here, $r_{s}=2M$ is the ``optical'' event horizon, while $r_{ac-}=M(\xi-\sqrt{\xi^{2}-4\xi})$ and $r_{ac+}=M(\xi+\sqrt{\xi^{2}-4\xi})$ are the interior and exterior ``acoustic'' event horizons respectively \cite{Guo2020,Vieira2021}. The necessary condition for the existence of ``acoustic'' event horizon is $\xi \geq 4$. When tuning parameter satisfies $0 \le \xi < 4$, only ``optical'' horizon exists in acoustic Schwarzschild black hole, Both the interior and exterior ``acoustic'' event horizons disappear in this case. When tuning parameter $\xi=4$, the interior and exterior ``acoustic'' event horizons coincide with each other, and we get the extreme acoustic Schwarzschild black hole. When tuning parameter $\xi > 4$, there are three interesting regions in acoustic Schwarzschild black hole. In region $r<r_{s}$, both light rays (photons) and sound waves (phonons) cannot escape from the acoustic black hole. In region $r_{s}<r<r_{ac+}$, light rays could escape from acoustic black hole, while sound waves cannot. In region $r>r_{ac+}$, both light rays and sound waves could escape from the acoustic black hole. The reader could consult references \cite{Ge2019,Guo2020} for more discussions on ``optical'' and ``acoustic'' event horizons of acoustic Schwarzschild black hole.

\section{Gravitational Bending Effects for Acoustic Schwarzschild Black Hole: Results and Discussions \label{sec:4}}

In this section, the gravitational bending effect of acoustic Schwarzschild black hole is analyzed in details. Results and discussions on gravitational deflection angle, lens equation and Einstein ring are presented. In subsection \ref{sec:4a}, the gravitational deflection angle is calculated using Gauss-Bonnet theorem. In subsection \ref{sec:4b}, the gravitational deflection angle is obtained by solving the trajectory of null geodesics. Subsection \ref{sec:4c} presents results on weak gravitational lensing and Einstein ring of acoustic Schwarzschild black hole.

\subsection{Gravitational Deflection Angle for Acoustic Schwarzschild Black Hole --- Obtained Using Gauss-Bonnet Theorem \label{sec:4a}}

\begin{figure*}
	\includegraphics[width=0.625\textwidth]{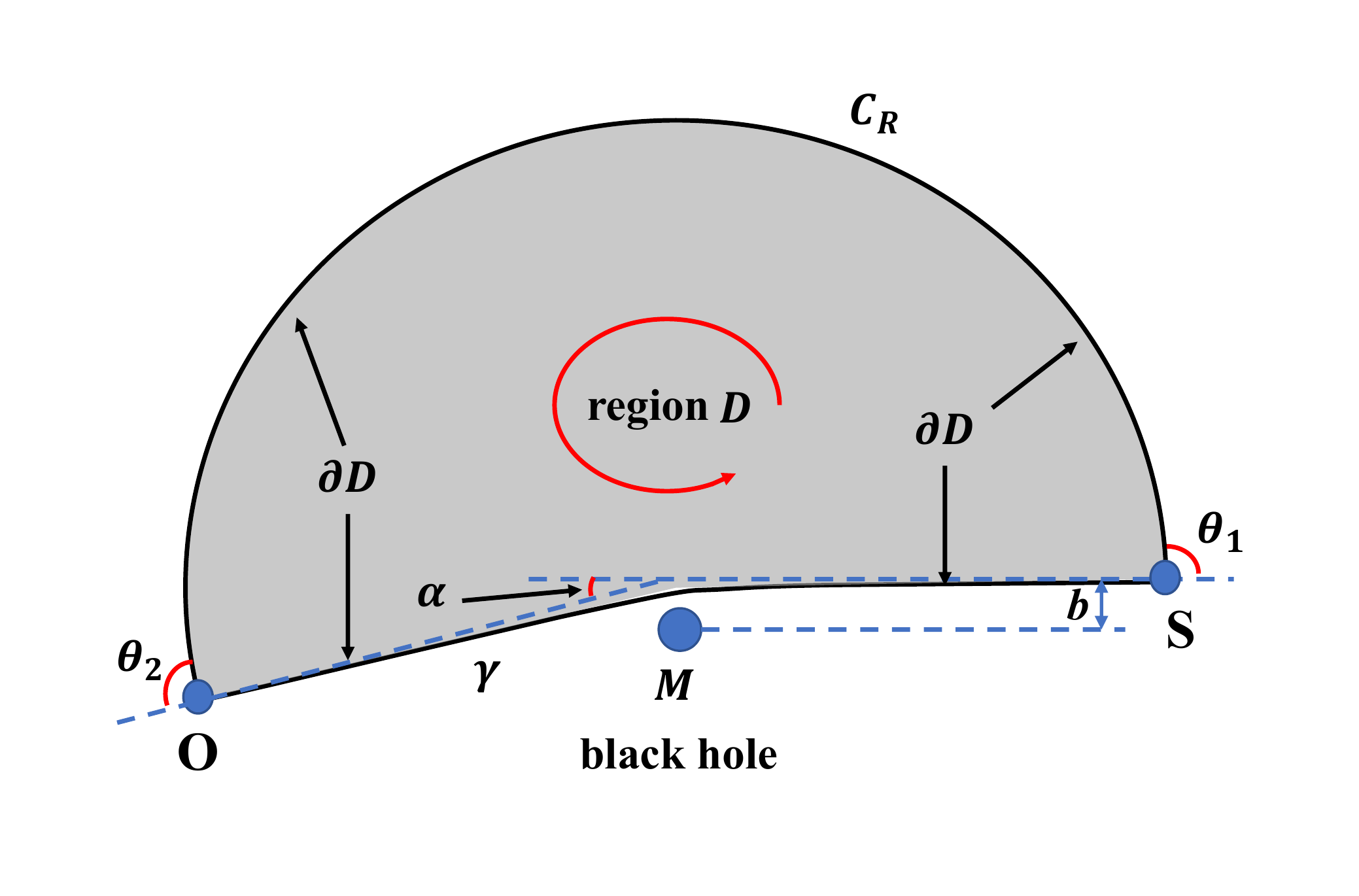}
	\caption{Calculating the gravitational deflection angle using the Gauss-Bonnet theorem. This figure shows the choice of region $D$ in the equatorial plane of optical manifold for asymptotically flat spacetime, which makes the application of Gauss-Bonnet theorem in equation (\ref{Gauss-Bonnet theorem}) available. In this figure, $O$ and $S$ label the locations of observer and particle source. The direction of boundary $\partial D$ in the contour integral $\int_{\partial D} \kappa_{g} dl$ is chosen to be counterclockwise. Note that for the particle orbit $\gamma$, this choice of direction (which is from $O$ to $S$) is opposite to the propagation of particles.}
	\label{figure1}
\end{figure*}

This subsection presents results and discussions on gravitational deflection angle. The gravitational deflection angle is obtained from Gauss-Bonnet theorem in cases where particle source (labeled by $S$) and observer (labeled by $O$) are both located at infinity. Moreover, the deflection angle obtained from Gauss-Bonnet theorem in cases where $O$ and $S$ are located at finite distance region is presented in Appendix \ref{appendix1}.

Gauss-Bonnet theorem is one of the most significant theorems in differential geometry and geometrical topology. It provide connections between geometry and topology in a curved manifold \cite{Chern}. In the Gauss-Bonnet theorem, topological properties of curved manifold, such as the Euler characteristic number, are reflected by pure geometric quantities (namely the various curvatures). Particularly, in a 2-dimensional curved manifold, the mathematical description of Gauss-Bonnet theorem is
\begin{equation}
	\int_{D} K dS + \int_{\partial D}\kappa_{g} dl + \sum_{i=1}^{N}\theta_{i} = 2 \pi \chi(D) 
	\label{Gauss-Bonnet theorem}
\end{equation}
Here, $D$ is a region in curved manifold, $K$ is the Gauss curvature, $\kappa_{g}$ is the geodesic curvature of boundary $\partial D$, $\chi(D)$ is the Euler characteristic number for region $D$, and $\theta_{i}$ is the exterior angle for each discontinuous point of boundary $\partial D$. In a regular and simply-connected region $D$ without singularities, its Euler characteristic number becomes $\chi(D)=1$.

In 2008, G. W. Gibbons and M. C. Werner developed an approach to calculate the gravitational deflection angle of light using Gauss-Bonnet theorem \cite{Gibbons2008}. In this approach, the gravitational deflection angle can be reflected by topological properties of curved spacetime, therefore it gives new insights on gravitational bending effects. Recently, this approach has been applied to many gravitational systems, and consistent results with traditional methods have been obtained \cite{Werner2012,Ishihara2016a,Ishihara2016b,Ovgun2018,Jusufi2018,Crisnejo2018a,Jusufi2018epjc,Javed2019epjc,Takizawa2020,Pantig2020}. 

In Gibbons and Werner's work, the gravitational deflection angle is calculated by applying the Gauss-Bonnet theorem in optical geometry / optical manifold. For a 4-dimensional asymptotically flat and spherically symmetric spacetime 
\begin{eqnarray}
	d\tau^{2} & = & g_{\mu\nu}dx^{\mu}dx^{\nu} \nonumber \\
	& = & f(r)dt^{2}
	-\frac{1}{f(r)}dr^{2}
	-r^{2}
	(d\theta^{2}+\sin^{2}\theta d\phi^{2}) \label{spacetime}
\end{eqnarray}
the optical geometry gives a 3-dimensional Riemannian manifold \cite{Gibbons2008,Werner2012,Abramowicz1988,Gibbons2009}
\begin{eqnarray}
	dt^{2} & = & g^{\text{OP}}_{ij}dx^{i}dx^{j} 
	= \frac{1}{f(r)}
	\bigg[
	\frac{1}{f(r)}dr^{2} 
	+ r^{2} \big( d\theta^{2}+\sin^{2}\theta d\phi^2 \big)
	\bigg] \nonumber \\
\end{eqnarray}
Here, the Greek indices $\mu$, $\nu$ run over 0,1,2,3; while the Latin indices $i$, $j$, $k$ run over 1,2,3, respectively. Actually, this optical geometry / optical metric is obtained by imposing the null constraint to the 4-dimensional spacetime metric. 
\begin{equation}
	\underbrace{d\tau^{2} = g_{\mu\nu}dx^{\mu}dx^{\nu}}_{\text{Spacetime Geometry}}
	\ \ \overset{d\tau^{2}=0}{\Longrightarrow} \ \ 
	\underbrace{dt^{2} = g^{\text{OP}}_{ij}dx^{i}dx^{j}}_{\text{Optical Geometry}}
\end{equation}
This 3-dimensional optical metric defines the so-called optical manifold or optical geometry in literature \cite{Abramowicz1988,Gibbons2009}. The propagation of light beams / sound waves, which is along null geodesic in original 4-dimensional spacetime manifold, still maintains geodesic in the optical manifold. Since the spacetime in equation (\ref{spacetime}) is spherically symmetric, without loss of generality, we can restrict the above optical manifold (as well as the optical metric $g^{\text{OP}}_{ij}$) in the equatorial plane 
\begin{equation}
	dt^{2}=\tilde{g}^{\text{OP}}_{ij}dx^{i}dx^{j}
	=\frac{1}{[f(r)]^{2}}dr^{2}+\frac{r^2}{f(r)}d\phi^{2}
	\label{optical metric1}
\end{equation}
where polar angle is fixed at $\theta=\pi/2$. In this way, a 2-dimensional Riemannian manifold is eventually obtained, and the application of Gauss-Bonnet theorem in equation (\ref{Gauss-Bonnet theorem}) becomes available.

The optical geometry (or optical manifold) is motivated by Fermat’s principle in optics. The classical Fermat’s principle states that: light rays always travel along particular spatial curves such that the optical distance $s_{ab}^{OP}=\int_{a}^{b} n(x)dx$ is minimal (where $n(x)$ is the reflective index). The classical Fermat’s principle is restricted in a flat space (or spacetime). However, this principle can be generalized into a stationary curved spacetime. In the stationary curved spacetime, one can have a global choice of stationary time $t$, massless particles (which travel along null geodesics in spacetime) starting from a fixed emission point at a given stationary time $t_{a}$ would eventually make the stationary arrival time $t_{b}$ minimal, which means the variational condition $\delta\big[\int_{a}^{b}dt\big]=0$ is satisfied. In this way, the particle orbits must be spatial geodesics in the optical geometry. In the optical geometry, the optical metric $dt^{2}=g_{ij}^{OP}dx^{i}dx^{j}$ gives the infinitesimal change of stationary time square, and the spatial geodesics in optical geometry always make the variation $\delta\big[\int_{a}^{b}dt\big]=0$. In the generalization of Fermat's principle to stationary curved spacetime, the spatial length in optical geometry $l_{ab}^{OP}=\int_{a}^{b}dt=\int_{a}^{b}\sqrt{g_{ij}^{OP}dx^{i}dx^{j}}$ effectively plays the role of “optical distance”. The discussion of the generalized Fermat's principle in a static or stationary curved spacetime can be found in references \cite{Werner2012,Jusufi2018,Gibbons2009,Landau1975,Perlick2000,LinQ2008}.

In the calculations of gravitational deflection angles, the region $D$ in Gauss-Bonnet theorem is chosen in the following way. For an asymptotically flat spacetime, when $O$ and $S$ are both very far from the central massive black hole, $D$ is a simply connected region in equatorial plane such that $O$ and $S$ are connected by its boundary $\partial D$. The boundary $\partial D$ is generated by two parts: the particle orbit $\gamma$ from $S$ to $O$, and a circular arc curve $C_{R}$ connected with $O$ and $S$. The central black hole is located outside the region $D$, so that spacetime singularities are excluded from $D$. This region is schematically illustrated in figure \ref{figure1}. 

For the picked region $D$ in figure \ref{figure1}, the exterior angles for discontinuous points of boundary $\partial D$ are $\theta_{1} \approx \theta_{2} \approx \pi/2$. Therefore, in the Gauss-Bonnet theorem, we have
\begin{equation}
	\sum_{i=1}^{N}\theta_{i} = \theta_{1} + \theta_{2} \approx \pi \label{exterior angle}
\end{equation}
Further, the Euler characteristic number for the simply connected region $D$ in figure \ref{figure1} is
\begin{equation}
	\chi(D)=1 \label{Eular characteristic number}
\end{equation}

The boundary $\partial D$ in figure \ref{figure1} consists of two parts: the particle orbit $\gamma$ and the outer circular arc $C_{R}$. If $O$ and $S$ are both very far from the central massive black hole, we can simply use the limit $R \to \infty$. Because the particle orbit $\gamma$ in gravitational field is a geodesic curve in optical geometry, its geodesic curvature $\kappa_{g}(\gamma)$ vanishes. Then the contour integral reduces to  
\begin{eqnarray} 
	\int_{\partial D}\kappa_{g} dl & = & \int_{\gamma}\kappa_{g}(\gamma) dl 
	+\lim_{R \to \infty} \int_{C_{R}}\kappa_{g}(C_{R}) dl \nonumber
	\\
	& = & \lim_{R \to \infty} \int_{C_{R}}\kappa_{g}(C_{R}) dl 
\end{eqnarray}
For asymptotically flat spacetimes, many studies have shown that the integration of geodesic curvature $\kappa_{g}$ along circular arc $C_{R}$ in $R \to \infty$ limit reduces to \cite{Javed2019,Jusufi2021}
\begin{equation} 
	\int_{\partial D}\kappa_{g} dl = \lim_{R \to \infty} \int_{C_{R}}\kappa_{g}(C_{R}) dl  
	\approx \pi+\alpha
	\label{geodesic curvature contour integral}
\end{equation}
where $\alpha$ is the gravitational deflection angle of particles traveling along null geodesics. For this kind of spacetime, the geodesic curvature of outer circular arc $C_{R}$ in equatorial plane approaches to $\kappa_{g}(C_{R}) \to 1/R + O(1/R^{2})$ when $R \to \infty$.

The spacetime generated by acoustic Schwarzschild black hole is a asymptotically flat spacetime. The contour integral of geodesic curvature follows the result in equation (\ref{geodesic curvature contour integral}). This can be directly confirmed using the optical geometry of acoustic Schwarzschild black hole. For outer circular arc $C_{R}$, its geodesic curvature can be calculated using the geodesic curvature formula for constant radius curve \cite{ChernWH}
\begin{eqnarray}
	\kappa_{g}(C_{r}) & = & \frac{1}{2\sqrt{\tilde{g}^{\text{OP}}_{rr}}} 
	\frac{\partial \ln \tilde{g}^{\text{OP}}_{\phi\phi}}{\partial r} \nonumber
	\\
	& = & \frac{f(r)}{r}-\frac{1}{2}\frac{\partial f(r)}{\partial r} \nonumber
	\\
	& = & \frac{1}{r}-\frac{3M(1+\xi)}{r^{2}}+\frac{16M^{2}\xi}{r^{3}}
	-\frac{20M^{3}\xi}{r^{4}}  \label{geodesic curvature CR}
\end{eqnarray}
For acoustic Schwarzschild black hole, the function $f(r)$ defined in equation (\ref{spacetime metric2}) has been used.
From this result, the integration of geodesic curvature $\kappa_{g}$ along the boundary $\partial D$ becomes
\begin{eqnarray}
	\int_{\partial D}\kappa_{g} dl & = & \lim_{R \to \infty} \int_{C_{R}}\kappa_{g}(C_{R}) dl \nonumber
	\\
	& = & \lim_{R \to \infty} \int_{\phi_{\text{source}}}^{\phi_{\text{observer}}}\kappa_{g}(C_{R}) R d\phi \nonumber
	\\ 
	& \approx & \lim_{R \to \infty}
	\int_{0}^{\pi+\alpha} 
	\bigg[ 
	\frac{1}{R}-\frac{3M(1+\xi)}{R^{2}} \nonumber
	\\
	&        &  \ \ \ \ \ \ \ \ \ \ \ \ \ \ \ \ 
	+\frac{16M^{2}\xi}{R^{3}} 
	-\frac{20M^{3}\xi}{R^{4}}
	\bigg] R d\phi \nonumber
	\\
	& = & \pi+\alpha \label{contour integral}
\end{eqnarray}
Because $O$ and $S$ are both located at infinity, so the approximations $\phi_{\text{source}} \approx 0$ and $\phi_{\text{observer}} \approx \pi + \alpha$ can be used, as indicated in figure \ref{figure1}. Note that this result is consistent with equation (\ref{geodesic curvature contour integral}) obtained in many literatures \cite{Javed2019,Jusufi2021}. As explained in figure \ref{figure1}, the direction of boundary $\partial D$ in the contour integral $\int_{\partial D} \kappa_{g} dl$ in equation (\ref{contour integral}) is chosen to be counterclockwise. For a particle orbit $\gamma$, this choice of direction (which is from $O$ to $S$) is opposite to the propagation of particles. 

Combining the exterior angle in equation (\ref{exterior angle}), geodesic curvature in equation (\ref{geodesic curvature contour integral}) and Euler characteristic number in equation (\ref{Eular characteristic number}), the Gauss-Bonnet theorem in equation (\ref{Gauss-Bonnet theorem}) eventually leads to
\begin{eqnarray}
	&   & \int_{D} K dS + \int_{\partial D}\kappa_{g} dl + \sum_{i=1}^{N}\theta_{i}
	= 2 \pi \chi(D) \nonumber
	\\
	& \Rightarrow & \int_{D} K dS+ (\pi+\alpha) + \pi 
	= 2 \pi
\end{eqnarray}
Therefore, the gravitational deflection angle of particle traveling along null geodesics can be calculated through the integral
\begin{eqnarray}
	\alpha & = & -\int_{D} K dS \nonumber
	\\
	& = &  -\int_{\phi_{\text{source}}}^{\phi_{\text{observer}}} d\phi 
	\int_{r(\gamma)}^{\infty} K \frac{r}{[f(r)]^{3/2}}dr
	\label{Gauss-Bonnet deflection}
\end{eqnarray} 
This is the gravitational deflection angle in spherical symmetric and asymptotically flat spacetime. In many acoustic gravity models, particles traveling along null geodesics of acoustic black holes maybe quasiparticles and phonons (sound-waves) \cite{Visser2005,Visser1998}. Furthermore, it should be pointed out that, the above results correspond to the cases where $O$ and $S$ are both from infinity (namely both $O$ and $S$ are very far from the central massive black hole). When $O$ and $S$ are located at finite distance region, the algorithm is presented in the Appendix \ref{appendix1}.

In the equatorial plane of optical manifold, the Gauss curvature in optical manifold can be calculated through the following expression in Riemannian geometry \cite{Chern,ChernWH}
\begin{eqnarray}
	K & = & -\frac{1}{\sqrt{\tilde{g}^{\text{OP}}}} 
	\bigg[ 
	\partial_{\phi} 
	\bigg(
	\frac{\partial_{\phi}\big(\sqrt{\tilde{g}^{\text{OP}}_{rr}}\big)}{\sqrt{\tilde{g}^{\text{OP}}_{\phi\phi}}}
	\bigg) 
	+\partial_{r} 
	\bigg(
	\frac{\partial_{r}\big(\sqrt{\tilde{g}^{\text{OP}}_{\phi\phi}}\big)}{\sqrt{\tilde{g}^{\text{OP}}_{rr}}}
	\bigg) 
	\bigg] \nonumber
	\\
	& = & \frac{1}{2} f(r) \cdot \frac{d^{2}f(r)}{dr^{2}}
	- \bigg[ \frac{1}{2} \cdot \frac{df(r)}{dr} \bigg]^{2} \nonumber 
	\\
	& = & -\frac{2M(1+\xi)}{r^{3}}
	+\frac{3M^{2}(\xi^{2}+10\xi+1)}{r^{4}}
	-\frac{48M^{3}\xi(\xi+2)}{r^{5}} \nonumber
	\\
	&   &
	+\frac{8M^{4}\xi(27\xi+11)}{r^{6}}     
	-\frac{384M^{5}\xi^{2}}{r^{7}} 
	+\frac{240M^{6}\xi^{2}}{r^{8}} 
	\label{Gauss Curvature vacuum}        
\end{eqnarray}
and the surface area in the equatorial plane can be expressed as
\begin{eqnarray}
	dS & = &  \sqrt{\tilde{g}^{\text{OP}}} dr d\phi 
	=    \frac{r}{[f(r)]^{3/2}} dr d\phi   \nonumber
	\\
	& = &  \frac{rdrd\phi}
	{\bigg\{ \big( 1-\frac{2M}{r} \big)   \cdot \big[ 1-\xi\frac{2M}{r} \big( 1-\frac{2M}{r} \big) \big] \bigg\}^{3/2}}   \nonumber
	\\
	& = & \bigg[
	1+\frac{3M(1+\xi)}{r}
	+\frac{3M^{2}(5\xi^{2}+2\xi+5)}{2r^{2}}  \nonumber
	\\
	&   & \ \ 
	+\frac{M^{3}(35\xi^{3}-15\xi^{2}+9\xi+35)}{2r^{3}} \nonumber
	\\
	&   & \ \       
	+\frac{15M^{4}(21\xi^{4}-28\xi^{3}-2\xi^{2}+4\xi+21)}{8r^{4}} \nonumber
	\\
	&   &  \ \ 
	+\frac{3M^{5}(231\xi^{5}-525\xi^{4}+70\xi^{3}-10\xi^{2}+35\xi+231)}{8r^{5}} \nonumber
	\\
	&   & \ \ 
	+O\bigg(\frac{M^{6}}{r^{6}}\bigg)
	\bigg] rdrd\phi \label{surface area}    
\end{eqnarray}

To evaluate the integral in equation (\ref{Gauss-Bonnet deflection}), the radius of particle orbit $r(\gamma)=r(\phi)$ must be determined. In principle, the exact value of $r(\gamma)$ can be calculated by solving the trajectories of null geodesics. However, the purpose of Gibbons and Werner's approach is to calculate the deflection angle through basic calculus and topological properties, without solving any complicated differential equations. For this purpose, in actual calculations, we can use the leading order approximation for a particle orbit.
\begin{equation}
	r(\gamma)=r(\phi) \approx \frac{b}{\sin\phi}
\end{equation}
Here, $b$ is the impact parameter indicated in figure \ref{figure1}. This is exactly the orbit in Newtonian gravity. Furthermore, we also assume that the gravitational deflection angle is not large, which correspond to the weak gravitational lensing. In this case, when $O$ and $S$ are located at infinity, we have the approximations
\begin{equation}
	\phi_{\text{source}} \approx 0; \ \ \ 
	\phi_{\text{observer}} \approx \pi+\alpha \approx \pi
\end{equation}
With the $r(\gamma)$, $\phi_{\text{source}}$ and $\phi_{\text{observer}}$ obtained, the integration of Gauss curvature can be evaluated.
\begin{eqnarray}
	\alpha & = & -\int_{D} K dS \nonumber
	\\
	& = & -\int_{\phi_{\text{source}}}^{\phi_{\text{observer}}} d\phi 
	\int_{r(\gamma)}^{\infty} K \frac{r}{[f(r)]^{3/2}} dr \nonumber
	\\
	& \approx & -\int_{0}^{\pi} d\phi 
	\int_{\frac{b}{\sin{\phi}}}^{\infty} K \frac{r}{[f(r)]^{3/2}} dr   \nonumber
	\\
	& = &  -\int_{0}^{\pi} d\phi
	\int_{\frac{b}{\sin{\phi}}}^{\infty} 
	\bigg[ 
	-\frac{2M(1+\xi)}{r^{2}} - \frac{3M^{2}(\xi^{2}-6\xi+1)}{r^{3}} \nonumber
	\\
	&   & \ \ \ \ \ \ \ \ \ \ \ \ \ \ \ \ \ \ \ \ \ \ \ \ 
	-\frac{6M^{3}(\xi-1)(\xi^{2}-4\xi-1)}{r^{4}}   \nonumber
	\\
	&   & \ \ \ \ \ \ \ \ \ \ \ \ \ \ \ \ \ \ \ \ \ \ \ \  
	+O\bigg(\frac{M^{5}}{r^{5}}\bigg)
	\bigg] dr \nonumber
	\\
	& = &  \int_{0}^{\pi} 
	\bigg[
	\frac{2M(1+\xi)\sin{\phi}}{b} 
	+\frac{3M^{2}(\xi^{2}-6\xi+1)\sin^{2}\phi}{2b^{2}} \nonumber
	\\
	&   & \ \ \ \ \ \ \ 
	+\frac{2M^{3}(\xi-1)(\xi^{2}-4\xi-1)\sin^{3}\phi}{b^{3}} 
	+ O\bigg(\frac{M^{4}}{b^{4}}\bigg)     
	\bigg] d\phi \nonumber
	\\
	& =  &  \frac{4M(1+\xi)}{b} 
	+\frac{3M^{2}(\xi^{2}-6\xi+1)\pi}{4b^{2}} \nonumber
	\\
	&   & 
	+\frac{8M^{3}(\xi-1)(\xi^{2}-4\xi-1)}{3b^{3}}  
	+O\bigg(\frac{M^{4}}{b^{4}}\bigg)  
	\label{deflection angle vacuum}
\end{eqnarray}


\begin{figure}
	\includegraphics[width=0.525\textwidth]{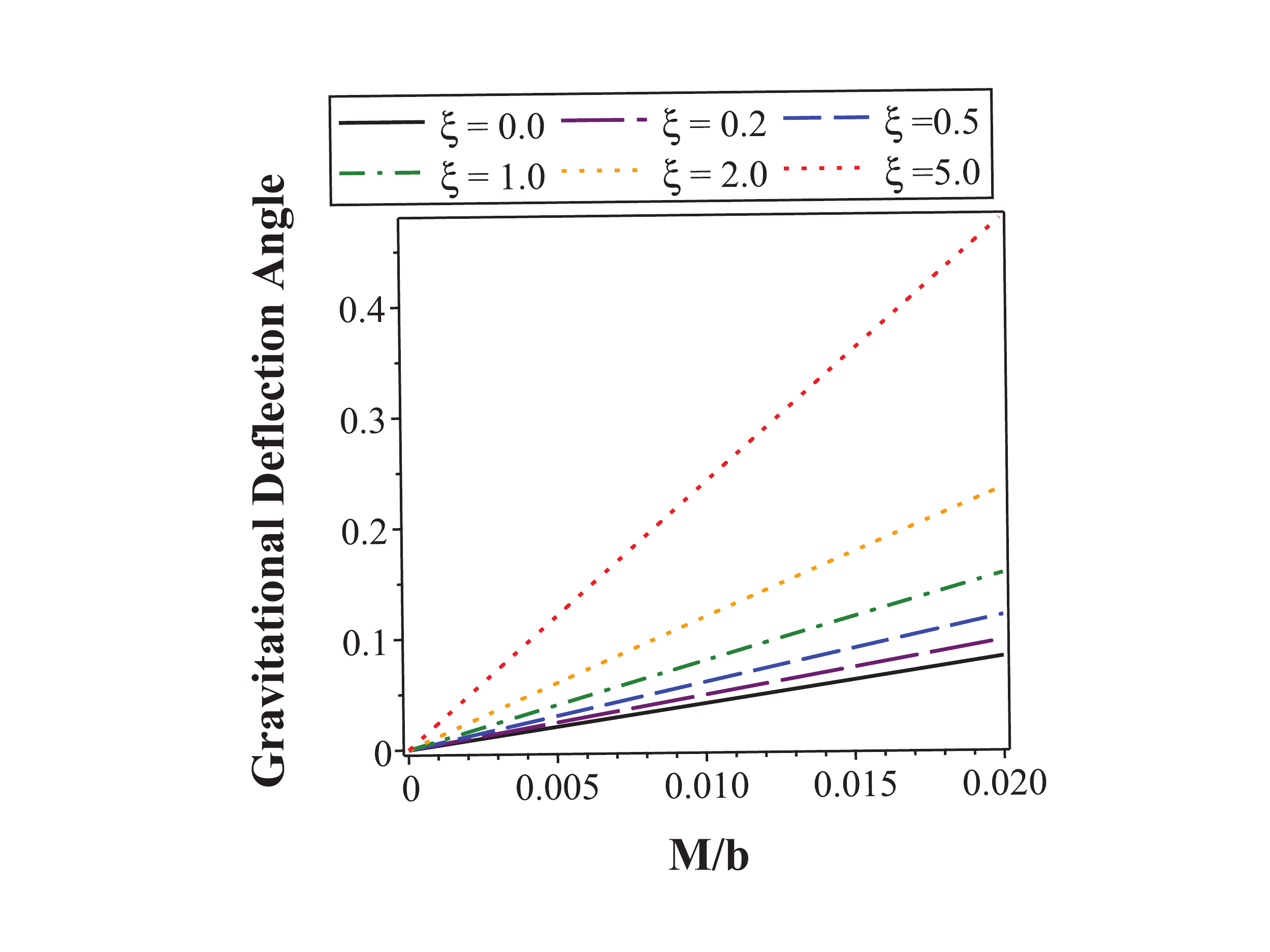}
	\caption{The gravitational deflection angles in acoustic Schwarzschild black hole for several $\xi$ values. The horizontal axis labels the $M/b$ and the vertical axis labels the gravitational deflection angle $\alpha$. In this figure, the gravitational deflection angles are plotted in region $0 \le M/b \le 0.02$, which correspond to the cases of weak gravitational lensing.} 
	\label{figure2}
\end{figure}

The above result in equation (\ref{deflection angle vacuum}) is the gravitational deflection angle for acoustic Schwarzschild black hole in cases where $O$ and $S$ are both located at infinity. When the distances between $O$, $S$ and central black hole are finite, the gravitational deflection angle is presented in the Appendix \ref{appendix1}. Detailed calculations  confirmed that the result in equation (\ref{deflection angle vacuum}) is consistent with the gravitational deflection angle for finite distance (which is given in equation (\ref{geodesic curvature finite distance})). Furthermore, when tuning parameter $\xi=0$, equation (\ref{deflection angle vacuum}) exactly reduces to the gravitational deflection angle for conventional Schwarzschild spacetime. Recall that the leading order contribution of gravitational deflection angle for Schwarzschild spacetime is 
\begin{equation}
	\alpha_{\text{Schwarzschild}}=\frac{4M}{b}
	+O\bigg(\frac{M^{2}}{b^{2}}\bigg)
\end{equation}
It can demonstrate the validity of the Gauss-Bonnet approach as well as the calculations in the present work \cite{footnote}.

However, when tuning parameter $\xi \neq 0$, deviations between conventional Schwarzschild black hole and acoustic Schwarzschild black hole emerge, and the gravitational bending behavior may exhibit different features. In figure \ref{figure2}, the gravitational deflection angles of particles traveling along null geodesics in acoustic Schwarzschild black hole are presented for several $\xi$ values. Since the expression of gravitational deflection angle is expressed in power series of $M/b$, we mostly focus on the region where $M/b$ is sufficiently small. These cases correspond to the weak gravitational lensing observations. Figure \ref{figure2} displays the gravitational deflection angle in region $0 \le M/b \le 0.02$. From this figure, it is indicated that the deflection angles for acoustic Schwarzschild black hole are larger than that for conventional Schwarzschild black hole. Moreover, when tuning parameter $\xi$ increases, the deflection angle $\alpha$ enlarges. This is because that the deflection angle is mostly dominant by the first term $4M(1+\xi)/b$ in equation (\ref{deflection angle vacuum}). Therefore, acoustic Schwarzschild black hole with large tuning parameter could greatly intensify the gravitational bending effect. Black holes with larger gravitational deflection angle are easier to catch and observe in gravitational lensing observations. The acoustic black holes, with the presence of moving fluids and sound-waves, may be more easily detectable through gravitational bending effects and gravitational lensing observations.

To see the strong gravitational lensing cases, we need to plot the gravitational deflection angle for larger $M/b$. And we would expect the emergence of non-linearity for deflection angle due to the higher order power series of $M/b$. However, in region where $M/b$ is large (which is $M/b > 0.1$ and will be shown in the next subsection), the gravitational deflection angle would become vary large. In these cases, the approximation $\phi_{\text{observer}} \approx \pi + \alpha \approx \pi$ used in the integration of Gauss curvature no longer satisfied. Therefore, in the strong gravitational lensing cases, the gravitational deflection angle obtained using Gauss-Bonnet theorem in equation (\ref{deflection angle vacuum}) would produce non-negligible discrepancies, and more precise methods are needed to explore the gravitational bending effect in this region. In the next subsection, we calculate the gravitational deflection angle for acoustic Schwarzschild black hole by solving the trajectory of null geodesics.

\subsection{Gravitational Deflection Angle for Acoustic Schwarzschild Black Hole --- Obtained by Solving the Trajectory of Null Geodesics \label{sec:4b}}

This subsection presents the gravitational deflection angle for acoustic Schwarzschild black hole based on an alternative method. It is the geodesic method, in which the gravitational deflection angle is calculated by solving the trajectory of null geodesics. After developed for many years, this approach has been tested by large numbers of observations, and it has become a widely adopted approach in physics and astronomy. This approach can duel with both weak and strong gravitational lensings. Historically, many famous problems on gravitational bending and gravitational lensing can be solved using this geodesic method \cite{Virbhadra2000,Weinberg1972,Virbhadra1998b,Eiroa2004,Keeton2005,Iyer2007,Virbhadra2009,Kim2021}. In this subsection, we follow the procedure in reference \cite{Weinberg1972,Virbhadra1998b} and obtain the gravitational deflection angle of acoustic Schwarzschild black hole through the trajectories of null geodesics.

The key point of this geodesic method is calculating the variation of azimuthal angle $\phi$ along the trajectory of null geodesics. For a spherically symmetric spacetime
\begin{equation}
	d\tau^{2} = B(r)dt^{2} -A(r)dr^{2} 
	-r^{2}(d\theta^{2}+\sin^{2}\theta d\phi^{2})
	\label{spacetime metric 3}
\end{equation}
the geodesic equations restricted in the equatorial plane ($\theta=\pi/2$) can be expressed as
\begin{subequations}
	\begin{eqnarray}
		\frac{d^{2}t}{d\lambda^{2}} 
		+ \frac{B'(r)}{B(r)} \frac{dt}{d\lambda} \frac{dr}{d\lambda} & = & 0  \nonumber
		\\
		\\
		\frac{d^{2}r}{d\lambda^{2}}
		+ \frac{B'(r)}{2A(r)} \bigg( \frac{dt}{d\lambda} \bigg)^{2}
		+ \frac{A'(r)}{2A(r)} \bigg( \frac{dr}{d\lambda} \bigg)^{2}  
		- \frac{r}{A(r)} \bigg( \frac{d\phi}{d\lambda} \bigg)^{2} & = & 0  \nonumber
		\\
		\\
		\frac{d^{2}\phi}{d\lambda^{2}}
		+ \frac{2}{r} \frac{d\phi}{d\lambda} \frac{dr}{d\lambda} & = & 0  \nonumber
		\\
	\end{eqnarray}
\end{subequations}
where $\lambda$ is the affine parameter, and $B'(r)=dB/dr$. From the above geodesic equations, two conserved quantities can be obtained \cite{Weinberg1972}
\begin{eqnarray}
	J & \equiv & r^{2}\frac{d\phi}{d\lambda} = \text{constant}
	\\
	-E & \equiv & A(r) \bigg( \frac{dr}{d\lambda} \bigg)^{2}
	+\frac{J^{2}}{r^{2}}-\frac{1}{B(r)} = \text{constant}
\end{eqnarray}
Here, $J$ is the conserved angular momentum of test particles. For particles traveling along timelike and null geodesics, the value of $E$ becomes
\begin{eqnarray}
	\bigg\{
	\begin{aligned}
		E=0 & & \text{particles traveling along null geodesics} &  \\
		E>0 & & \text{particles traveling along timelike geodesics} & 
	\end{aligned} 
\end{eqnarray}
In the Newtonian limit (when particles move slowly in a weak gravitational field), for massive particles traveling along timelike geodesics, $\epsilon/2 = (1-E)/2$ is the conserved energy per unit mass \cite{footnote2}. Using these conserved quantities, the differential equation of particle orbit $r(\gamma)$ can be obtained
\begin{equation}
	\frac{A(r)}{r^{4}} \bigg( \frac{dr}{d\phi} \bigg)^{2}
	+ \frac{1}{r^{2}} -\frac{1}{J^{2}\cdot B(r)}
	= - \frac{E}{J^{2}} \label{orbital equation 1}
\end{equation}

When particle reaches its minimal distance $r_{0}$ to the central black holes, the derivative $d\phi/dr=0$. Then the conserved angular momentum can be expressed as
\begin{equation}
	J = r_{0}\sqrt{\frac{1}{B(r_{0})}-E} 
	= \frac{r_{0}}{\sqrt{B(r_{0})}} 
	\label{conserved angular momentum}
\end{equation}
where $E=0$ has been used for particles traveling along null geodesics. Therefore, for particles traveling along null geodesics, the change of radius with respect to azimuthal angle is derived from equation (\ref{orbital equation 1}) and equation (\ref{conserved angular momentum})
\begin{eqnarray}
	\frac{dr}{d\phi} & = & \pm 
	\frac{\sqrt{A(r)}}{\sqrt{r^{2}\big[\frac{1}{J^{2}B(r)}-\frac{E}{J^{2}}-\frac{1}{r^{2}}\big]}} 
	=   \pm \frac{\sqrt{A(r)}}{\sqrt{r^{2}\big[\frac{1}{J^{2}B(r)}-\frac{1}{r^{2}}\big]}} \nonumber
	\\
	& = & \pm \frac{\sqrt{A(r)}}{\sqrt{\big(\frac{r}{r_{0}}\big)^{2}\cdot\frac{B(r_{0})}{B(r)}-1}}
\end{eqnarray}
The plus and minus signs are determined in the following way. Along the trajectory of null geodesics, as particle moves from infinity to the minimal distance $r_0$, the azimuthal angle $\phi$ increases as radius $r$ decreases. However, as particle moves from the minimal distance $r=r_0$ to infinity $r=\infty$, the azimuthal angle $\phi$ increases as radius $r$ increases. In this way, we have
\begin{eqnarray}
	\frac{dr}{d\phi} & = & -\frac{\sqrt{A(r)}}{\sqrt{\big(\frac{r}{r_{0}}\big)^{2}\cdot\frac{B(r_{0})}{B(r)}-1}} < 0 \ \ \ \text{from $r=\infty$ to $r=r_{0}$}  \nonumber 
	\\
	\frac{dr}{d\phi} & = & \frac{\sqrt{A(r)}}{\sqrt{\big(\frac{r}{r_{0}}\big)^{2}\cdot\frac{B(r_{0})}{B(r)}-1}} > 0 \ \ \ \text{from $r=r_{0}$ to $r=\infty$}  \nonumber \\
\end{eqnarray}
When observer $O$ and particle source $S$ are both located at infinity, the gravitational deflection angle is just twice of the change $\Delta\phi$ as radius $r$ varies from $r=r_{0}$ to $r=\infty$ \cite{Weinberg1972,Virbhadra1998b}.
\begin{eqnarray}
	\alpha & = & 2|\phi(\infty)-\phi(r_{0})|-\pi \nonumber
	\\
	& = & 2\bigg| 
	\int_{r_{0}}^{\infty} \frac{d\phi}{dr}dr 
	\bigg|
	- \pi \nonumber
	\\
	& = & 2\int_{r_{0}}^{\infty}
	\frac{\sqrt{A(r)}}
	{r\sqrt{\big(\frac{r}{r_{0}}\big)^{2}\cdot\frac{B(r_{0})}{B(r)}-1}}
	dr -\pi 
	\label{gravitational deflection angle from null geodesic}
\end{eqnarray}
From this expression, it is evident that the gravitational deflection angle $\alpha$ increases monotonically as the minimal distance $r_{0}$ decreases.


Substituting the spacetime metric of acoustic Schwarzschild black hole in equation (\ref{spacetime metric}) into equation (\ref{gravitational deflection angle from null geodesic}), the gravitational deflection angle of particles traveling along null geodesics is calculated through the integral
\begin{eqnarray}
	\alpha & = & 2\int_{r_{0}}^{\infty}
	\frac{\sqrt{A(r)}}
	{r\sqrt{\big(\frac{r}{r_{0}}\big)^{2}\cdot\frac{B(r_{0})}{B(r)}-1}}
	dr -\pi \nonumber
	\\
	& = &  2\int_{r_{0}}^{\infty}
	\frac{dr}
	{r\sqrt{\big(\frac{r}{r_{0}}\big)^{2}\cdot f(r_{0})-f(r)}}
	-\pi \nonumber
	\\
	& = & 2\int_{r_{0}}^{\infty}
	\frac{dr}
	{r\sqrt{\big(\frac{r}{r_{0}}\big)^{2} \cdot \big[1-\frac{2M(1+\xi)}{r_{0}}+\frac{8\xi M^{2}}{r_{0}^{2}}-\frac{8\xi M^{3}}{r_{0}^{3}}\big]}} \nonumber
	\\
	&   & \ \ \ \ \ \ \ \ 
	\frac{}{\overline{-\big[1-\frac{2M(1+\xi)}{r}+\frac{8\xi M^{2}}{r^{2}}-\frac{8\xi M^{3}}{r^{3}}\big]}} 
	\ -\pi 
\end{eqnarray}
To simplify the numerical calculations, one can introduce the following dimensionless parameters 
\begin{equation}
	x=\frac{r}{2M} \ \ \ \ 
	x_{0}=\frac{r_{0}}{2M} \ \ \ \  z=\frac{x_{0}}{x}=\frac{r_{0}}{r}  
\end{equation}
Then the gravitational deflection angle can be expressed as
\begin{eqnarray}
	\alpha & = & 2\int_{x_{0}}^{\infty}
	\frac{dx}
	{x\sqrt{\big(\frac{x}{x_{0}}\big)^{2}\cdot\big[1-\frac{1+\xi}{x_{0}}+\frac{2\xi}{x_{0}^{2}}-\frac{\xi}{x_{0}^{3}}\big]}}
	\\
	&   & \ \ \ \ \ \ \ \ 
	\frac{}{\overline{-\big[1-\frac{1+\xi}{x}+\frac{2\xi}{x^{2}}-\frac{\xi}{x^{3}}\big]}} 
	-\pi \nonumber
	\\
	& = & 2\int_{0}^{1}
	\frac{dz}{\sqrt{\big[1-\frac{1+\xi}{x_{0}}+\frac{2\xi}{x_{0}^{2}}-\frac{\xi}{x_{0}^{3}}\big]}} \nonumber
	\\
	&   & \ \ \ \ \ \ \ \ \frac{}{\overline{-z^{2}\cdot\big[1-\frac{1+\xi}{x_{0}}\cdot z +\frac{2\xi}{x_{0}^{2}}\cdot z^{2} -\frac{\xi}{x_{0}^{3}}\cdot z^{3}\big]}} 
	-\pi \nonumber
	\\
	& = & 2\int_{0}^{1}
	\frac{dz}{\sqrt{P_{0}(x_{0})-P(z)}} -\pi 
	\label{gravitational deflection angle --- hyper-elipitical integral}
\end{eqnarray}
Here, $P_{0}(x_{0})$ is a constant depends on $x_{0}$, $P(z)$ is a polynomial function of fifth order 
\begin{eqnarray}
	P(z) & \equiv & z^{2} - \frac{1+\xi}{x_{0}} z^{3}
	+ \frac{2\xi}{x_{0}^2} z^{4} 
	- \frac{\xi}{x_{0}^{3}} z^{5}
	\\
	P_{0}(x_{0}) & \equiv & 1 - \frac{1+\xi}{x_{0}}
	+ \frac{2\xi}{x_{0}^{2}}
	- \frac{\xi}{x_{0}^{3}}
\end{eqnarray}

\begin{figure*}
	\includegraphics[width=0.495\textwidth]{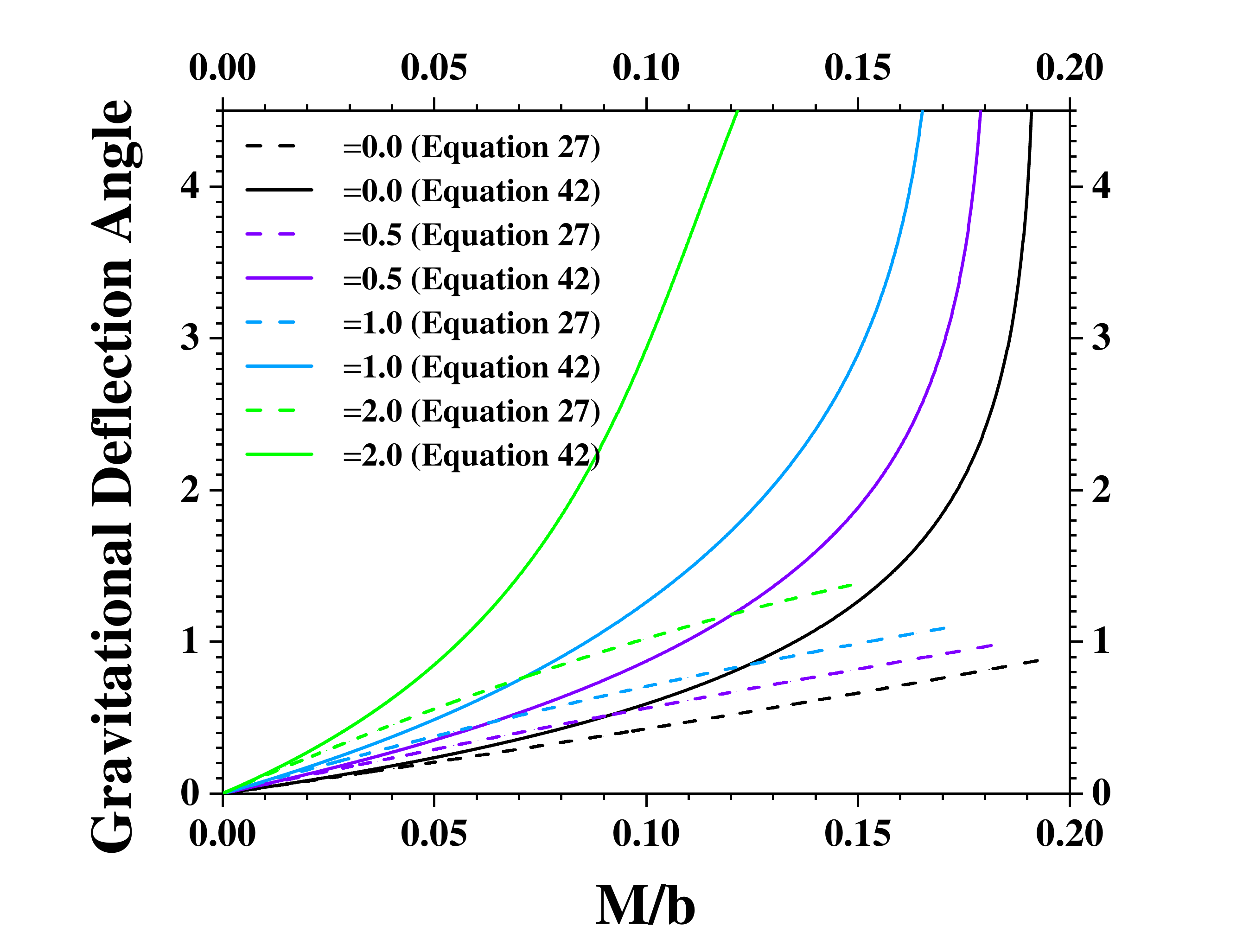}
	\includegraphics[width=0.495\textwidth]{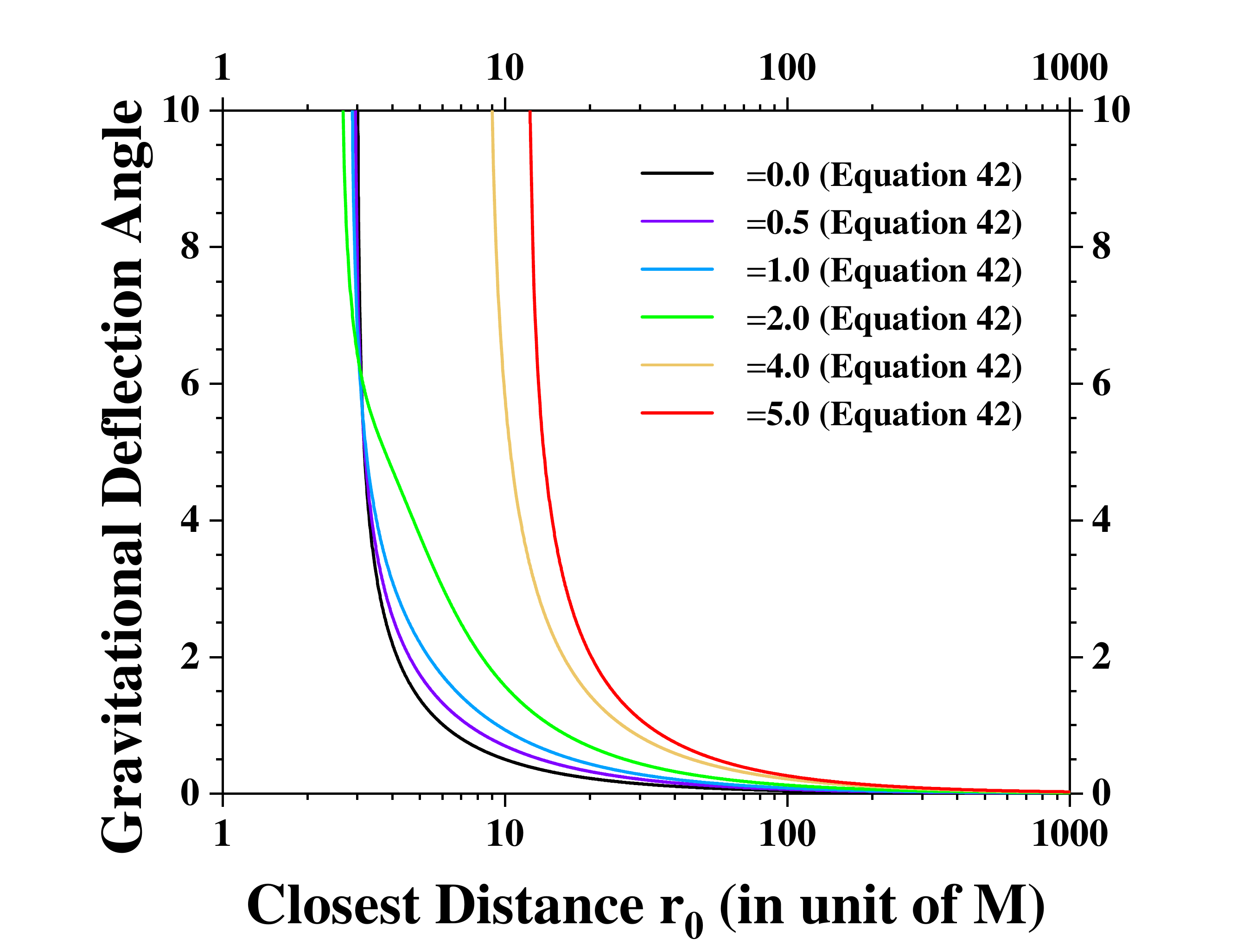}
	\caption{The gravitational deflection angles for acoustic Schwarzschild black hole obtained by solving the trajectories of null geodesics. We have chosen the cases $\xi=0$, $\xi=0.5$, $\xi=1$, $\xi=2$, $\xi=4$ and $\xi=5$ as representative examples. In the left panel, the gravitational deflection angles are plotted in region $0 \le M/b \le 0.2$. Comparative results of deflection angles obtained using Gauss-Bonnet theorem in equation (\ref{deflection angle vacuum}) and hyper-elliptical integral in equation (\ref{gravitational deflection angle --- hyper-elipitical integral}) are displayed. In right panel, the gravitational deflection angles are plotted with respective to the minimal distance $r_{0}$ in the particle orbit $r(\gamma)$. Since the gravitational deflection angle obtained from Gauss-Bonnet theorem is a power series of $M/b$ rather than $1/x_{0}=2M/r_{0}$, so we only plot the deflection angle obtained from the hyper-elliptical integral in equation (\ref{gravitational deflection angle --- hyper-elipitical integral}) in the right panel.}
	\label{figure3}
\end{figure*}

In this way, the gravitational deflection angle can be calculated using a hyper-elliptic integral
\begin{equation}
	F=\int_{z_{1}}^{z_{2}}\frac{dz}{\sqrt{a-P_{n}(z)}}
\end{equation}
where $a$ is a constant, $P_{n}(z)$ is a polynomial of $z$ in $n$-th order. This hyper-elliptic integral can be expanded into power series analytically \cite{Loiseau}, but the procedure is complicated, which is left in the Appendix \ref{appendix2}. Instead, in this subsection, we employ a numerical scheme to calculate this hyper-elliptic integral.

In the numerical calculation of gravitational deflection angle, the dimensionless parameter $x_{0}=r_{0}/2M$ in hyper-elliptic integral is determined by solving the minimal distance $r_{0}$ to central black holes in the trajectory of null geodesic $r(\gamma)=r(\phi)$. Specially, the minimal distance $r_{0}$ can be calculated using the conserved quantities $J$, $E$ and the impact parameter $b$. Following the notations in reference \cite{Weinberg1972}, the impact parameter is defined as 
\begin{equation}
	b \equiv \frac{J}{\epsilon} = \frac{J}{1-E} 
	= \frac{r_{0}}{\sqrt{B(r_{0})}} = \frac{r_{0}}{\sqrt{f(r_{0})}} 
\end{equation}
which leads to
\begin{equation}
	\frac{2M}{b} = \frac{\sqrt{f(r_{0})}}{x_{0}}
	= \frac{\sqrt{1-\frac{1+\xi}{x_{0}}+\frac{2\xi}{x_{0}^{2}}-\frac{\xi}{x_{0}^3}}}{x_{0}} 
\end{equation}
where we have used equation (\ref{conserved angular momentum}) and $E=0$ for particles traveling along null geodesics.  

The numerical results of gravitational deflection angle for acoustic Schwarzschild black hole obtained using the above hyper-elliptic integral are presented in figure \ref{figure3}. In the left panel, comparative results between the two approaches (one is the Gauss-Bonnet theorem, the other is the geodesic method) are plotted in region $0 \le M/b \le 0.2$. In the right panel, the gravitational deflection angles calculated using the geodesic method and the hyper-elliptic integral are plotted with respective to the minimal distance $r_{0}$ in particle orbit $r(\gamma)$. From this figure, it is clearly indicated that, when tuning parameter $\xi$ is larger, the deflection angle $\alpha$ also increases. Therefore, the gravitational bending effect for acoustic Schwarzschild black hole is greatly enhanced, compared with the conventional Schwarzschild black hole. In the weak gravitational lensing cases when $M/b$ is very small, gravitational deflection angles calculated from these two approaches converge to each other.

\begin{table*}
	\caption{Shadow and circular photon orbit for acoustic Schwarzschild black hole. In this table, the radius of black hole shadow and unstable circular orbit are listed for several tuning parameter $\xi$. The $R_{sh}=b_{\text{cr}}^{\text{max}}$ is the shadow radius detected by observer located at infinity, and $R_{co}=r_{0}^{\text{min}}$ is radius of unstable circular orbit for acoustic Schwarzschild black hole. As we have explained in section \ref{sec:3}, when tuning parameter $0<\xi<4$, only the ``optical'' horizon $r_{h}=2M$ exists. When tuning parameter $\xi>4$, the additional interior and exterior ``acoustic'' event horizons $r_{ac-}=M(\xi-\sqrt{\xi^{2}-4\xi})$ and $r_{ac+}=M(\xi+\sqrt{\xi^{2}-4\xi})$ emerge. Furthermore, in this work, we only discuss the gravitational deflection of particles traveling along the null geodesics outside the ``optical'' and ``acoustic'' event horizons ($r>r_{ac+}$ and $r>r_{s}$).}
	\label{table0}
	\vspace{2mm}
	\begin{tabular}{lccc}
		\hline
		& Tuning Parameter $\xi$ & Shadow Radius $R_{sh}=b_{\text{cr}}^{\text{max}}$ & Radius of Unstable Circular Orbit $R_{co}=r_{0}^{\text{min}}$
		\\
		\hline
		& $\xi=0$   & $5.196M \approx 3\sqrt{3}M$ & $3.001M\approx 3M$
		\\
		& $\xi=0.2$ & $5.315M$ & $2.978M$
		\\
		& $\xi=0.5$ & $5.508M$ & $2.936M$
		\\
		& $\xi=1.0$ & $5.869M$ & $2.851M$
		\\
		& $\xi=2.0$ & $6.739M$ & $2.613M$
		\\
		& $\xi=4.0$ & $18.351M$ & $8.705M$
		\\
		& $\xi=5.0$ & $23.780M$ & $12.058M$
		\\
		\hline
	\end{tabular}
\end{table*}

However, when $M/b$ is large, the gravitational deflection angles obtained from the two approaches exhibit notable differences. There are two reasons lead to this discrepancy. Firstly, the deflection angle in equation (\ref{deflection angle vacuum}) obtained using Gauss-Bonnet theorem is a third-order expansion of $M/b$. Higher-order contributions, which is non-negligible when $M/b$ is larger, are not included. Secondly, even in the second and third order contributions of $M/b$, gravitational deflection angles from these two approaches still do not agree with each other. This is probably duel to the approximations $\phi_{\text{observer}} \approx \pi + \alpha \approx \pi$ and $r(\gamma)=r(\phi) \approx b/\sin\phi$ used in the integration of Gauss curvature in subsection \ref{sec:4a}. This issue is discussed in details in Appendix \ref{appendix2}. 

In the Appendix \ref{appendix2}, we expand the gravitational deflection angle $\alpha$ in equation (\ref{gravitational deflection angle --- hyper-elipitical integral}), which is expressed using a hyper-elliptic integral, into two kinds of power series. One is the power series of $1/x_{0}$, the other is the power series of $M/b$. Firstly, the expansion of gravitational deflection angle in equation (\ref{gravitational deflection angle --- hyper-elipitical integral}) over $1/x_{0}$ gives
\begin{eqnarray}
	\alpha & = & \frac{2(1+\xi)}{x_{0}}
	+ \bigg[ \bigg( \frac{15\pi}{16}-1 \bigg) \cdot (1+\xi)^{2} - \frac{3\pi}{2} \cdot \xi \bigg] \cdot \frac{1}{x_{0}^{2}} \nonumber
	\\
	&   & + \bigg[ 
	(1+\xi)^{3}\cdot\bigg(\frac{61}{12}-\frac{15\pi}{16}\bigg)
	+ \xi(1+\xi)\cdot\bigg(\frac{3\pi}{2}-14\bigg) \nonumber
	\\
	&   & \ \ \ 
	+ \xi\cdot\frac{8}{3}  
	\bigg] \cdot \frac{1}{x_{0}^{3}} 
	+ O\bigg( \frac{1}{x_{0}} \bigg)^{4} 
\end{eqnarray}
When the tuning parameter $\xi=0$, this power series agree with the gravitational deflection angle for conventional Schwarzschild spacetime \cite{Virbhadra2000}
\begin{eqnarray}
	\alpha_{\text{Schwarzschild}} 
	& = & \frac{2}{x_{0}}
	+ \bigg( \frac{15\pi}{16}-1 \bigg) \cdot \frac{1}{x_{0}^{2}}
	+ O\bigg( \frac{1}{x_{0}^{3}} \bigg) \nonumber
	\\
	& = & \frac{4M}{r_{0}}
	+ \bigg( \frac{15\pi}{16}-1 \bigg) \cdot \frac{4M^{2}}{r_{0}^{2}}
	+ O\bigg( \frac{8M^{3}}{r_{0}^{3}} \bigg) 
\end{eqnarray}
Secondly, the power expansion of gravitational deflection angle over $M/b$ gives
\begin{eqnarray}
	\alpha & = & \frac{4M(1+\xi)}{b}
	+\frac{3\pi M^{2}(5\xi^{2}+2\xi+5)}{4b^{2}} \nonumber
	\\
	&   & +\frac{8M^{3}(16\xi^{3}+8\xi+16)}{3b^{3}}
	+O\bigg(\frac{M^{4}}{b^{4}}\bigg) 
	\label{deflection angle --- power series00}
\end{eqnarray}
From this equation, it is clearly indicated that the power series obtained from Gauss-Bonnet theorem agree with the power series obtained from the hyper-elliptic integral only in the lower-order contributions of $M/b$. There are non-negligible discrepancies in higher-order contributions of $M/b$. So the gravitational deflection angles obtained from Gauss-Bonnet theorem and the geodesic method could exhibit notable discrepancies when $M/b$ is large, as we have plotted in figure \ref{figure3}. 

From figure \ref{figure3}, it is also indicated that, in the strong gravitational lensing when $M/b$ is large and $r_{0}\to R_{co}$ (where $R_{co}$ is the radius of circular photon orbit), the gravitational deflection angle increases rapidly as $b$ and $r_{0}$ varies. In some critical points $b_{\text{cr}}^{\text{max}}$ and $r_{0}^{\text{min}}$, the gravitational deflection angle could become divergent. The critical point $b_{\text{cr}}^{\text{max}}$ would correspond to the black hole shadow radius detected by observer located at infinity, and the critical point $r_{0}^{\text{min}}$ would correspond to the unstable circular orbit \cite{Perlick2022,Ling2021}. The numerical results on unstable circular orbit radius $R_{co}=r_{0}^{\text{min}}$ and black hole shadow radius $R_{sh}=b_{\text{cr}}^{\text{max}}$ are displayed in table \ref{table0} for acoustic Schwarzschild black hole. It is also worth noting that the circular photon orbit / photon sphere for acoustic Schwarzschild black hole may exhibit rather different behaviors as tuning parameter $\xi$ changes. When tuning parameter satisfies $0<\xi<3.32$, there is only one circular photon orbit, which is unstable. When $\xi>3.32$, the acoustic Schwarzschild black hole may have three circular photon orbits (two unstable ones and a stable one in between). The circular photon orbit $R_{co}=r_{0}^{\text{min}}$ given in table \ref{table0} is just the outermost unstable circular photon orbit. 
From this table, it can be clearly seen that: when tuning parameter $\xi$ increases, the radius of black hole shadow $R_{sh}=b_{\text{cr}}^{\text{max}}$ (which is detected by observer located at infinity) increases. Combining the results in figure \ref{figure3} and table \ref{table0}, we can draw the conclusion: the acoustic black holes, with the presence of moving fluids and sound-waves, have larger gravitational deflection angle and black hole shadow radius. They may be more easily detectable through gravitational bending effects and gravitational lensing observations.

In subsection \ref{sec:4a} and \ref{sec:4b}, we have obtained the gravitational deflection angle for acoustic Schwarzschild black hole within two approaches. One is based on the Gauss-Bonnet theorem in geometrical topology, the other is the geodesic method (by solving the trajectory of null geodesics). Results from these two approaches are consistent with each other in the weak gravitational lensing cases where $M/b$ is small. However, in the strong gravitational lensing cases where $M/b$ is large, there are non-negligible discrepancies. The results obtained from Gauss-Bonnet theorem in equation (\ref{deflection angle vacuum}) is suitable only in the weak gravitational lensing. In the strong gravitational lensing cases, some approximations used in the integration of Gauss curvature no longer valid, and the results based on the Gauss-Bonnet theorem could underestimate the gravitational deflection angle.

\subsection{Weak Gravitational Lensing and Einstein Ring \label{sec:4c}}

\begin{figure*}
	\includegraphics[width=0.85\textwidth]{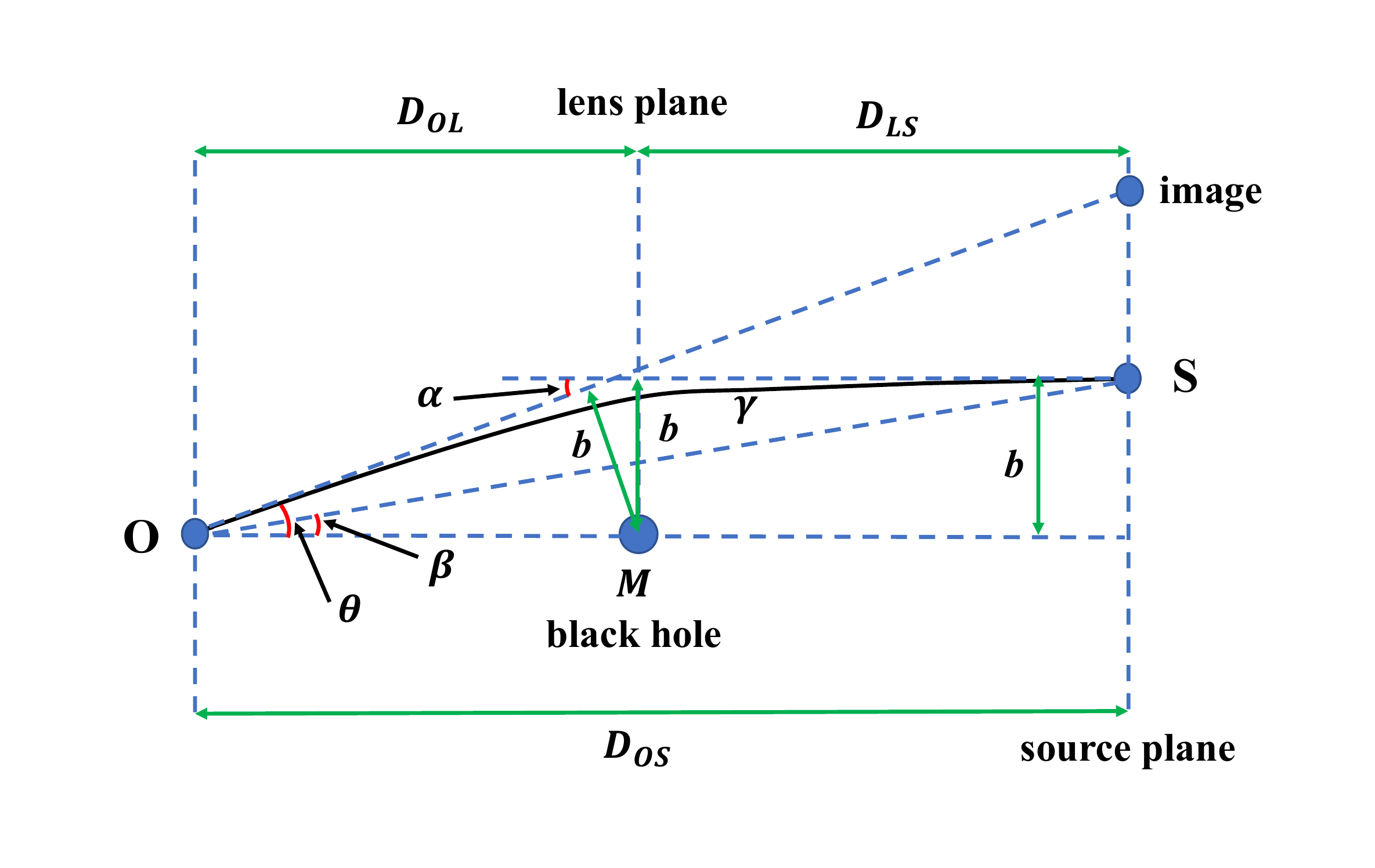}
	\caption{The schematically plot of the weak gravitation lensing. The particles emitted from a distant source (labeled by $S$) are lensed by a central black hole. In this figure, $D_{\text{OS}}$ is the distance between the observer (labeled by $O$) and source plane, $D_{\text{OL}}$ is the distance between $O$ and lens plane, $D_{\text{LS}}$ is the distance between lens plane and source plane. The central black hole is in the middle of lens plane. The angle $\beta$ denotes the angular position of $S$ with respect to the ``optical axis'' $\text{OM}$, and $\theta$ is the angular position of the lensed image of $S$ detected by $O$. The $\alpha$ is the gravitational deflection angle of particles traveling along null geodesics, and $b$ is the impact parameter.}
	\label{Weak gravitational lensing figure}
\end{figure*}

As introduced in section \ref{sec:1}, acoustic black holes can be generated in many ways, not only in condensed matter systems, but also from astrophysical black hole systems with fluid surrounded. Particularly, the relativistic and transonic accretion onto astrophysical black holes provide examples of acoustic black holes \cite{Das2004,Das2006}. The transonic accretion of superfluid dark matter on black hole systems also provide a scenario to realize analogue black holes \cite{Berezhiani2015,Ge2019}. If acoustic Schwarzschild black holes exist in our universe, they could be detected and observed from gravitational lensing observations. 

In gravitational lensing observations, physical observables are mostly constrained by lens equation, which is given by \cite{Bozza2008,Islam2020,Jusufi2021b}
\begin{equation}
	D_{\text{OS}}\tan\beta = \frac{D_{\text{OL}}\sin\theta-D_{\text{LS}}\sin(\alpha-\theta)}{\cos(\alpha-\theta)} \label{lens equation}
\end{equation} 
Here, $D_{\text{OS}}$ is the distance between the observer $O$ and source plane, $D_{\text{OL}}$ is the distance between $O$ and lens plane, $D_{\text{LS}}$ is the distance between lens plane and source plane, the angles $\beta$ and $\theta$ are indicated in figure \ref{Weak gravitational lensing figure}. It is worth noting that the angle $\theta$ in lens equation is different from the angle $\theta_{i}$ in Gauss-Bonnet theorem. In Gauss-Bonnet theorem, $\theta_{i}$ is the exterior angle for each discontinuous point of boundary $\partial D$. However, the angle $\theta$ in equation (\ref{lens equation}) is the angular position of the lensed image of $S$ detected by $O$. The positions and images of the lensed objects can be achieved by solving the lens equation (\ref{lens equation}). In the weak gravitational lensing, for distant $O$ and $S$, we have the approximations $\tan\beta \approx \beta$, $\sin\theta \approx \theta$, $\sin(\alpha-\theta) \approx \alpha-\theta$ and $\cos(\alpha-\theta) \approx 1$. Then the lens equation (\ref{lens equation}) reduces to \cite{Bozza2001,Petters2001,Mollerach2002}
\begin{equation}
	\beta = \theta - \frac{D_{\text{LS}}}{D_{\text{OS}}} \cdot \alpha
\end{equation} 	
The relation $D_{\text{OS}} = D_{\text{OL}}+D_{\text{LS}}$ has been used to derive this equation. The angular radius of Einstein ring is calculated by taking $\beta = 0$. In the weak gravitational lensing where $M/b$ is small, the gravitational deflection angle obtained from Gauss-Bonnet theorem in equation (\ref{deflection angle vacuum}) can be applied appropriately. In this way, angular radius of Einstein ring is given by
\begin{eqnarray}
	\theta_{\text{E}} & = & \frac{D_{\text{LS}}}{D_{\text{OS}}} \cdot \alpha \nonumber
	\\
	& = & \frac{D_{\text{LS}}}{D_{\text{OS}}} \cdot
	\bigg[
	\frac{4M(1+\xi)}{b} 
	+\frac{3M^{2}(\xi^{2}-6\xi+1)\pi}{4b^{2}} \nonumber
	\\
	&   &   +\frac{8M^{3}(\xi-1)(\xi^{2}-4\xi-1)}{3b^{3}} 
	+O\bigg(\frac{M^{4}}{b^{4}}\bigg)
	\bigg]
	\label{Einstein ring}
\end{eqnarray}
Further, in the weak gravitational lensing case, the impact parameter $b$ satisfies
\begin{equation}
	b \approx D_{\text{OL}}\sin\theta_{\text{E}} \approx D_{\text{OL}}\theta_{\text{E}}
	\label{impact parameter approximation}
\end{equation}
Combing equation (\ref{Einstein ring}) and equation (\ref{impact parameter approximation}), the angular radius of Einstein ring $\theta_{\text{E}}$ can be solved.

\begin{table}
	\caption{Einstein ring of acoustic Schwarzschild black hole. The angular radii $\theta_{\text{E}}$ of Einstein ring are presented for several tuning parameter $\xi$. In this table, the mass of acoustic Schwarzschild black hole is set as $M = 4.3 \times 10^{6}M_{\odot}$, and the distances between observer $O$, lens plane and source plane are chosen to be $D_{\text{LS}}=D_{\text{OS}}=8.3\ \text{kpc}$. In this table, we have excluded the unphysical solutions of Einstein ring $\theta_{\text{E}}$ (namely when $\theta_{\text{E}}$ gets complex values or negative real values).}
	\label{table1}
	\vspace{2mm}
	\begin{tabular}{lccc}
		\hline
		& Tuning Parameter & Angular radius of Einstein ring &
		\\
		\hline
		& $\xi=0$   & $\theta_{\text{E}}=1.45\ \text{arcsec}$ &
		\\
		& $\xi=0.2$ & $\theta_{\text{E}}=1.59\ \text{arcsec}$ &
		\\
		& $\xi=0.5$ & $\theta_{\text{E}}=1.78\ \text{arcsec}$ &
		\\
		& $\xi=1.0$ & $\theta_{\text{E}}=2.05\ \text{arcsec}$ &
		\\
		& $\xi=2.0$ & $\theta_{\text{E}}=2.51\ \text{arcsec}$ &
		\\
		& $\xi=5.0$ & $\theta_{\text{E}}=3.55\ \text{arcsec}$ &
		\\
		\hline
	\end{tabular}
	\vspace{2mm}
\end{table}

Although acoustic black holes could be generated by mechanisms in condensed matter physics, high-energy physics and cosmology. However, the realization of acoustic black holes only reported in condensed matter systems. So far, the astrophysical observational constrains on acoustic Schwarzschild black hole mass $M$ and tuning parameter $\xi$ are extremely weak. In this way, we can freely choose $M = 4.3 \times 10^{6}M_{\odot}$ and $D_{\text{LS}}=D_{\text{OL}}=8.3\ \text{kpc}$, which are the mass and distance for Sgr A*, as the representative example to shown the Einstein ring of astrophysical acoustic Schwarzschild black hole. To see the influence of tuning parameter on weak gravitational lensing, the angular radii of Einstein rings are calculated for several tuning parameters $\xi$, and numerical results are presented in table \ref{table1}. From this table, it is notified that the Einstein ring of acoustic Schwarzschild black hole is larger that of conventional Schwarzschild black hole. For larger tuning parameter $\xi$, the size of Einstein ring magnifies. Therefore, the acoustic Schwarzschild black holes, with the presence of moving fluids and sound-waves, could present larger Einstein ring images than the conventional Schwarzschild black hole, which make them more easily detectable in weak gravitational lensing observations.

\section{Summary and Prospects \label{sec:5}}

In this work, we study the gravitational bending effect of acoustic black hole. This category of black holes attracted broad research interests in recent years, for they open-up new directions for theoretical and experimental / observational explorations of black holes. Some kinds of acoustic black holes could offer connections between dynamics of black holes and tabletop experiments in laboratories. The investigations on acoustic black holes may be beneficial for experimental / observational explorations of black holes as well as theoretical investigations on analogue gravity models. To choose a simple and typical example, we consider the acoustic Schwarzschild black hole in this work. The acoustic Schwarzschild black hole could reflect some universal properties of more complex acoustic black holes, similar to the conventional Schwarzschild black hole in general relativity. In this work, the gravitational deflection angle of particles traveling along null geodesics, weak gravitation lensing and Einstein ring of acoustic Schwarzschild black hole are calculated and discussed in details.

In the calculation of gravitational deflection angle, two approaches have been applied. One is through the Gauss-Bonnet theorem in geometrical topology, the other is the geodesic method. In the Gauss-Bonnet approach, the gravitational deflection angle  is calculated from the integration of Gauss curvature in the equatorial plane of optical manifold. The gravitational bending effect of black holes is directly connected with the geometrical and topological properties of the optical geometry (the geodesic curvature $\kappa_{g}$, Gauss curvature $K$, and the Euler characteristic number $\chi(D)$). In the geodesic method, the gravitational deflection angle is obtained by solving the trajectories of null geodesics. In the weak gravitational lensing cases where $M/b$ is sufficiently small, the results obtained using two approaches agree with each other. However, in the strong gravitational cases where $M/b$ is not small, there are non non-negligible discrepancies between results obtained using the two approaches. Detailed analysis shows that the results obtained from Gauss-Bonnet approach is suitable only in the weak gravitational lensing. In the strong gravitational lensing cases, some approximations used in the integration of Gauss curvature are not satisfied, and the results based on the Gauss-Bonnet approach underestimate the gravitational deflection angle.

When tuning parameter $\xi=0$, our results reduce to the gravitational deflection angle in conventional Schwarzschild black hole correctly. 
However, when tuning parameter is nonzero, the gravitational bending in acoustic Schwarzschild black hole is greatly different from that in conventional Schw-arzschild black hole. Numerical calculations show that the gravitational deflection angle increases when tuning parameter $\xi$ becomes larger. From these results, we can conclude that, compared with the conventional Schwarzschild black hole, the gravitational bending effect for acoustic Schwarzschild black hole is greatly enhanced.

In the weak gravitational lensing of acoustic Schwarzschild black hole, the angular radius of Einstein ring is solved through equation
$ \theta_{\text{E}} = \alpha \cdot D_{\text{LS}}/D_{\text{OS}} $.
Numerical results in this work indicate that the Einstein ring of acoustic Schwarzschild black hole is larger that of conventional Schwarzschild black hole. Further, the size of Einstein ring magnifies when tuning parameter $\xi$ increases. The acoustic black holes, with the presence of moving fluids and sound-waves, may be more easily detectable in gravitational bending effects and weak gravitational lensing observations.

The approaches adopted in this work are general and model independent. It can be applied to arbitrary asymptotically flat acoustic black hole with spherical symmetry. In future studies, it is necessary to investigate the gravitational bending of more complex acoustic black holes, such as the acoustic Reissner-Nordstr\"om black hole \cite{Ling2021}, or the charged acoustic black hole with nonlinear electrodynamics. It is also interesting to extend our work to the gravitational bending of rotating acoustic black holes, such as the slowly rotating acoustic black holes proposed by H. S. Vieira \emph{et al.} \cite{Vieira2021b}, through which the influences from angular momentum of particles in acoustic black holes can be effectively analyzed. Furthermore, the gravitational bending and gravitational lensing of low-dimensional acoustic black holes in condensed matter systems can also be effectively analyzed using the similar approach. We wish that the present work could shed some light on the physics of black holes, as well as the analogue gravity models in high-energy physics and condensed matter physics. 

As we have introduced in section \ref{sec:1}, the realization of acoustic black holes in astrophysical systems are mostly through the relativistic and transonic accretion onto central black holes, or black holes in the bath of cosmological microwave background. The accumulating observations and high-precision measurements for black holes in the center of galaxies with an accretion disk, such as the Sgr A* in our galaxy, would possibly provide opportunities to search the astrophysical acoustic black holes. The numerical results and predictions in the present work may be test from these observations / measurements in future.

\begin{acknowledgments}
We acknowledge helpful discussions with Peng Wang, Xing-Da Liu and Ming Li. The authors should thank to the great efforts from all around the world during the pandemic period of COVID-19. This work is supported by the Natural Science Foundation of Chongqing (Grant No. cstc2020jcyj-msxmX0879),the Scientific Research Program of Chongqing Science and Technology Commission ("zhitongche program for doctors", Grant No. CSTB2022BSXM-JCX0100), the Scientific and Technological Research Program of Chongqing Municipal Education Commission (Grant No. KJQN202201126), and the Scientific Research Foundation of Chongqing University of Technology (Grants No. 2020ZDZ027).
\end{acknowledgments}

\appendix

\section{Gravitational Bending Effects for Acoustic Schwarzschild Black Hole ------ Particle Source and Observer Located at Finite Distance Region \label{appendix1}}  

In this appendix, we present results on gravitational bending of acoustic Schwarzschild black hole, with the finite distance effect taking into consideration. In this section, the locations of particle source and observer (which are labeled by $S$ and $O$ respectively) are much different from the cases in section \ref{sec:4}. Both $O$ and $S$ are not in the asymptotically flat region (the infinity). The distances between $O$, $S$ and central massive black hole are finite.

\begin{figure*}
	\includegraphics[width=0.665\textwidth]{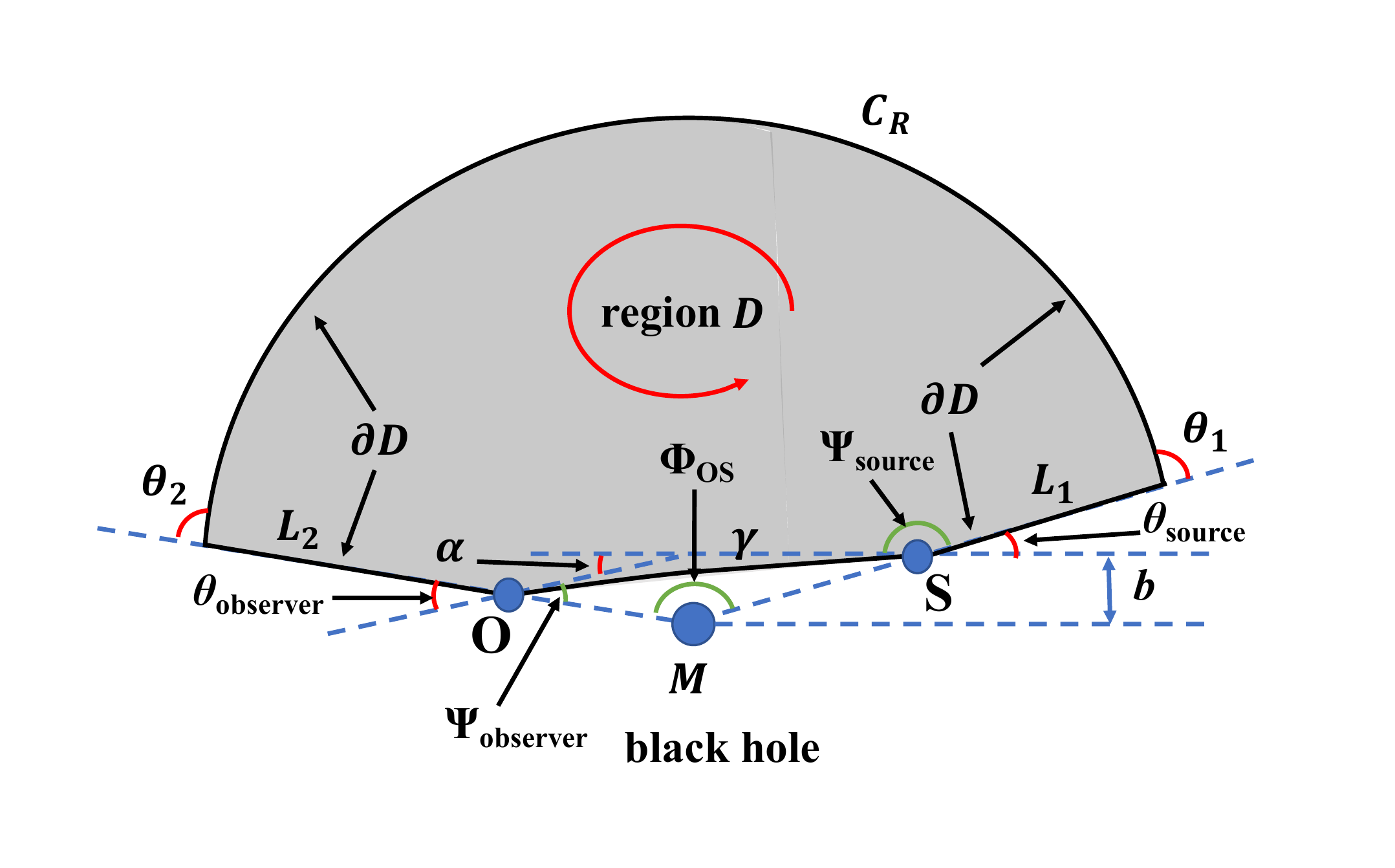}
	\caption{Calculating the gravitational deflection angle for finite distance using Gauss-Bonnet theorem. This figure shows the choice of region $D$ in the equatorial plane of optical manifold in cases where particle source (labeled by $S$) and observer (labeled by $O$) are both located at finite distance region. This region $D$ would make the application of Gauss-Bonnet theorem in equation (\ref{Gauss-Bonnet theorem}) available. The direction of boundary $\partial D$ in contour integral $\int_{\partial D} \kappa_{g} dl$ is chosen to be counterclockwise. Note that for the particle orbit $\gamma$, this choice of direction (which is from $O$ to $S$) is opposite to the propagation of particles. In this figure, $\Psi_{\text{source}}$ and $\Psi_{\text{observer}}$ are the angles between particle orbit $\gamma$ and the radial direction, which are exactly the same as the definition in Ishihara's work \cite{Ishihara2016a}. The angle $\phi_{\text{OS}}= \phi_{\text{observer}}-\phi_{\text{source}}$ is the change of azimuthal angle. After these definitions, the gravitational deflection angle for finite distance can be expressed as $\alpha_{\text{finite distance}} = \Psi_{\text{source}}-\Psi_{\text{observer}}+\phi_{\text{OS}}$.}
	\label{figure4}
\end{figure*}

To calculate the gravitational deflection angle for finite distance, we adopt the method developed by A. Ishihara \emph{et al.} \cite{Ishihara2016a,Ishihara2016b}. This method is an extension of the Gauss-Bonnet approach in subsection \ref{sec:4a}. In 2008, G. W. Gibbons and M. C. Werner first used Gauss-Bonnet theorem to calculate the gravitational deflection angle of black holes in the cases where $O$ and $S$ are located at infinity (the asymptotically flat region) \cite{Gibbons2008}. In 2016-2017, A. Ishihara \emph{et al.} extend this approach to the finite distance cases \cite{Ishihara2016a,Ishihara2016b}. In Ishihara's work, the gravitational deflection angle for finite distance is defined and calculated through
\begin{equation}
	\alpha_{\text{finite distance}} 
	=  \Psi_{\text{source}}-\Psi_{\text{observer}}+\phi_{\text{OS}}	
\end{equation}
Here, the angles $\Psi_{\text{source}}$ and $\Psi_{\text{observer}}$ are indicated in figure \ref{figure4}. The angle $\phi_{\text{OS}}$ is the change of azimuthal angle
\begin{equation}
	\phi_{\text{OS}} = \phi_{\text{observer}}-\phi_{\text{source}} 
\end{equation} 

In the calculation of gravitational deflection angle for finite distance, the region $D$ is chosen to be a simply connected region in the equatorial plane of optical manifold, which is illustrated in figure \ref{figure4}. The spacetime singularities are excluded from $D$. For this region, the exterior angles for discontinuous points of boundary $\partial D$ are: $\theta_{1}=\theta_{2}=\pi/2$, $\theta_{\text{observer}}=\Psi_{\text{observer}}$, $\theta_{\text{source}}=\pi-\Psi_{\text{source}}$. The sum of exterior angles in Gauss-Bonnet theorem becomes
\begin{eqnarray}
	\sum_{i=1}^{N}\theta_{i} 
	& = & \theta_{1} + \theta_{2} 
	+ \theta_{\text{observer}} + \theta_{\text{source}} \nonumber
	\\
	& = & 2\pi + \Psi_{\text{observer}} -\Psi_{\text{source}} \label{exterior angle finite distance}
\end{eqnarray}
Furthermore, for this simply connected region $D$, the Euler characteristic number is 
\begin{equation}
	\chi(D)=1 \label{Eular characteristic number2}
\end{equation}

The boundary $\partial D$ in figure \ref{figure4} consists of four parts: the particle orbit $\gamma$, the outer circular arc $C_{R}$, the lines $L_{1}$ and $L_{2}$. Here, $L_{1}$ and $L_{2}$ are lines with constant azimuthal angle $\phi$, and their geodesic curvature can be calculated through \cite{ChernWH}
\begin{equation}
	\kappa_{g}(L_{1}) = \kappa_{g}(L_{2}) 
	= -\frac{1}{2\sqrt{\tilde{g}^{\text{OP}}_{\phi\phi}}} 
	\frac{\partial \ln \tilde{g}^{\text{OP}}_{rr}}{\partial \phi} 
	=  0
\end{equation}
The particle orbit $\gamma$ in gravitational field is a geodesic curve in optical geometry, so its geodesic curvature $\kappa_{g}(\gamma)$ vanishes. Then the contour integral of geodesic curvature along the boundary $\partial D$ reduces to 
\begin{eqnarray} 
	\int_{\partial D}\kappa_{g} dl & = &   \int_{\gamma}\kappa_{g}(\gamma) dl 
	+\lim_{R \to \infty} \int_{C_{R}}\kappa_{g}(C_{R}) dl \nonumber
	\\
	&   & +\int_{L_{1}}\kappa_{g}(L_{1}) dl
	+\int_{L_{2}}\kappa_{g}(L_{2}) dl \nonumber
	\\
	& = & \lim_{R \to \infty} \int_{C_{R}}\kappa_{g}(C_{R}) dl 
\end{eqnarray}
For the circular arc curve $C_{R}$, the geodesic curvature has been given in equation (\ref{geodesic curvature CR}). Then the integration of geodesic curvature $\kappa_{g}$ along the boundary $\partial D$ becomes 	
\begin{eqnarray}
	\int_{\partial D}\kappa_{g} dl & = & 
	\lim_{R \to \infty} \int_{C_{R}}\kappa_{g}(C_{R}) dl \nonumber
	\\
	& =  & \lim_{R \to \infty}       \int_{\phi_{\text{source}}}^{\phi_{\text{observer}}} 
	\kappa_{g}(C_{R}) R d\phi \nonumber
	\\
	& = & \lim_{R \to \infty}
	\int_{\phi_{\text{source}}}^{\phi_{\text{observer}}} 
	\bigg[ 
	\frac{1}{R}-\frac{3M(1+\xi)}{R^{2}} \nonumber
	\\
	&   &   \ \ \ \ \ \ \ \ \ \ \ \ \ \ \ \ \ \ \ \   
	+\frac{16M^{2}\xi}{R^{3}} 
	-\frac{20M^{3}\xi}{R^{4}}
	\bigg] R d\phi \nonumber
	\\
	& = & \phi_{\text{observer}}-\phi_{\text{source}} \nonumber
	\\
	& = & \phi_{OS} \label{geodesic curvature finite distance}
\end{eqnarray}

Combining the exterior angles in equation (\ref{exterior angle finite distance}), contour integral of geodesic curvature in equation (\ref{geodesic curvature finite distance}) and Euler characteristic number in equation (\ref{Eular characteristic number2}), the Gauss-Bonnet theorem in equation (\ref{Gauss-Bonnet theorem}) eventually leads to
\begin{eqnarray}
	&   & \int_{D} K dS 
	+ \int_{\partial D}\kappa_{g} dl 
	+ \sum_{i=1}^{N}\theta_{i} \nonumber
	\\
	& = & \int_{D} K dS+ \phi_{OS} + (2\pi 
	+ \Psi_{\text{observer}} -\Psi_{\text{source}})  \nonumber
	\\
	& = & 2 \pi \chi(D) = 2 \pi
\end{eqnarray}
Therefore, in the asymptotically flat spacetime, the gravitational deflection angle can be calculated through
\begin{eqnarray}
	\alpha_{\text{finite distance}} & = &  \Psi_{\text{source}}-\Psi_{\text{observer}}+\phi_{\text{OS}} \nonumber
	\\
	& = & -\int_{D} K dS \nonumber 
	\\
	& = & -\int_{\phi_{\text{source}}}^{\phi_{\text{observer}}} d\phi 
	\int_{r(\gamma)}^{\infty} K \frac{r}{[f(r)]^{3/2}} dr  \nonumber
	\\ \label{integration finite}
\end{eqnarray}
This is the gravitational deflection angle of particles traveling along null geodesics in cases where $O$ and $S$ are both located at finite distance region.

\begin{widetext}
The same as in subsection \ref{sec:4a}, the radius of particle orbit $r(\gamma)=r(\phi)$ must be determined to evaluate the integration of Gauss curvature. To avoid solving any complicated differential equations for null geodesics, we can use the leading order perturbation for particle orbit, which is the orbit in Newtonian gravity. Namely, 
\begin{equation}
	r(\gamma)=r(\phi) \approx \frac{b}{\sin\phi} \label{photon orbit}
\end{equation}
Plug the Gauss curvature in equation (\ref{Gauss Curvature vacuum}), surface area in equation (\ref{surface area}) and the particle orbit in equation (\ref{photon orbit}) into the integration (\ref{integration finite}), one can finally obtain
\begin{eqnarray}
	\alpha_{\text{finite distance}} 
	& = & -\int_{D} K dS 
 	  =   -\int_{\phi_{\text{source}}}^{\phi_{\text{observer}}} d\phi 
	\int_{r(\gamma)}^{\infty} K \frac{r}{[f(r)]^{3/2}} dr \nonumber
	\\
	& \approx & -\int_{\phi_{\text{source}}}^{\phi_{\text{observer}}} d\phi 
	\int_{\frac{b}{\sin{\phi}}}^{\infty} K \frac{r}{[f(r)]^{3/2}} dr   \nonumber
	\\
	& = &  -\int_{\phi_{\text{source}}}^{\phi_{\text{observer}}} d\phi 
	\int_{\frac{b}{\sin{\phi}}}^{\infty} 
	\bigg[ 
	-\frac{2M(1+\xi)}{r^{2}} 
	-\frac{3M^{2}(\xi^{2}-6\xi+1)}{r^{3}} 
	-\frac{6M^{3}(\xi-1)(\xi^{2}-4\xi-1)}{r^{4}} 
	+O\bigg(\frac{M^{4}}{r^{5}}\bigg)
	\bigg] dr \nonumber
	\\
	& = &  \int_{\phi_{\text{source}}}^{\phi_{\text{observer}}} 
	\bigg[
	\frac{2M(1+\xi)\sin{\phi}}{b} 
	+\frac{3M^{2}(\xi^{2}-6\xi+1)\sin^{2}\phi}{2b^{2}} 
	+\frac{2M^{3}(\xi-1)(\xi^{2}-4\xi-1)\sin^{3}\phi}{b^{3}} 
	+O\bigg(\frac{M^{4}}{b^{4}}\bigg)     
	\bigg] d\phi \nonumber
	\\
	& = & \frac{2M(1+\xi)}{b} \cdot
	\bigg[   
	\cos(\phi_{\text{source}})-\cos(\phi_{\text{observer}}) \bigg] \nonumber
	\\
	&   & +\frac{3M^{2}(\xi^{2}-6\xi+1)}{4b^{2}} \cdot 
	\bigg[
	\cos(\phi_{\text{source}})\sin(\phi_{\text{source}})   -\cos(\phi_{\text{observer}})\sin(\phi_{\text{observer}})+\phi_{OS}
	\bigg] \nonumber
	\\
	&   &  +\frac{2M^{3}(\xi-1)(\xi^{2}-4\xi-1)}{3b^{3}} \cdot 
	\bigg\{
	\big[2+\sin^{2}(\phi_{\text{source}})\big] 
	\cdot \cos(\phi_{\text{source}}) 
	-\big[2+\sin^{2}(\phi_{\text{observer}})\big]
	\cdot \cos(\phi_{\text{observer}})
	\bigg\} \nonumber
	\\
	&   &  +O\bigg(\frac{M^{4}}{b^{4}}\bigg)
\end{eqnarray} 
To solve the azimuthal angles $\phi_{\text{source}}$ and $\phi_{\text{observer}}$ in a simpler way, we also employ the leading order solution for particle orbit
\begin{equation}
	r(\gamma) \approx \frac{b}{\sin\phi} \ \ 
	\Rightarrow \ \  
	\sin\phi \approx \frac{b}{r} = bu
\end{equation}
with $u=1/r$ to be the inverse of radial coordinate. Recall that the azimuthal angle in figure \ref{figure4} satisfies $\phi_{\text{source}}<\pi/2$ and $\phi_{\text{observer}}>\pi/2$. The trigonometric functions become
\begin{equation}
		\cos(\phi_{\text{source}}) = \sqrt{1-b^{2}u_{\text{source}}^{2}} 
		\ \ \ \ \ \ 
		\cos(\phi_{\text{observer}}) =  -\sqrt{1-b^{2}u_{\text{observer}}^{2}}
\end{equation}
Finally, the gravitational deflection angle of particles traveling along null geodesics can be expressed as
\begin{eqnarray}
	\alpha_{\text{finite distance}} & = & \frac{2M(1+\xi)}{b} \cdot
	\big[ 
	\sqrt{1-b^{2}u_{\text{source}}^{2}} 
	+\sqrt{1-b^{2}u_{\text{observer}}^{2}}
	\big]  	
	+ \frac{3M^{2}(\xi^{2}-6\xi+1)}{4b^{2}}  \cdot \nonumber
	\\
	&   & \ \ \ 
	\bigg[
	bu_{\text{source}} \sqrt{1-b^{2}u_{\text{source}}^{2}} 
	+ bu_{\text{observer}} \sqrt{1-b^{2}u_{\text{observer}}^{2}} 
	+\pi 
	-\arcsin(bu_{\text{observe}})
	-\arcsin(bu_{\text{source}})
	\bigg] \nonumber
	\\
	&   &  +\frac{2M^{3}(\xi-1)(\xi^{2}-4\xi-1)}{3b^{3}}
	\cdot 
	\bigg[  
	(2+b^{2}u_{\text{source}}^{2})  \sqrt{1-b^{2}u_{\text{source}}^{2}} 
	+(2+b^{2}u_{\text{observer}}^{2}) \sqrt{1-b^{2}u_{\text{observer}}^{2}} 
	\bigg] 
	+ O\bigg(\frac{M^{4}}{b^{4}}\bigg) \nonumber
	\\
	\label{Gauss-Bonnet deflection finite distance}
\end{eqnarray}
This is the gravitational deflection angle for acoustic Schwarzschild black hole where $O$ and $S$ are both located at finite distance region.

The gravitational deflection angle in acoustic Schwarzschild black hole spacetime for finite distance in equation (\ref{Gauss-Bonnet deflection finite distance}) is consistent with the results presented in subsection \ref{sec:4a}. When $O$ and $S$ are both located at infinity, which correspond to the case in subsection \ref{sec:4a}, we have the following limits
\begin{subequations}
	\begin{eqnarray} 
		r_{\text{source}} \to \infty \  & \Rightarrow & \  u_{\text{source}}=1/r_{\text{source}} \to 0  
		\\
		r_{\text{observer}} \to \infty \  & \Rightarrow & \  u_{\text{observer}}=1/r_{\text{observer}} \to 0 
	\end{eqnarray}
\end{subequations}
Plugging these relations into equation (\ref{Gauss-Bonnet deflection finite distance}), the gravitational deflection angle for acoustic Schwarzschild black hole becomes
\begin{equation}
	\alpha  \to  \frac{4M(1+\xi)}{b}
	+\frac{3M^{2}(\xi^{2}-6\xi+1)\pi}{4b^{2}} 
	+\frac{8M^{3}(\xi-1)(\xi^{2}-4\xi-1)}{3b^{3}} 
	+O\bigg(\frac{M^{4}}{b^{4}}\bigg) 
\end{equation}
In this case, the above results reduces to the gravitational deflection angle given in equation (\ref{deflection angle vacuum}).

\section{Analytical Expansion of the Gravitational Deflection Angle in Equation (\ref{gravitational deflection angle --- hyper-elipitical integral}) \label{appendix2}}

In this Appendix, we expand the gravitational deflection angle in equation (\ref{gravitational deflection angle --- hyper-elipitical integral}), which is related to a hyper-elliptical integral, into power series. Two kinds of power series are presented in this appendix. One is the power series of $1/x_{0}=2M/r_{0}$, the other is the power series of $M/b$. From these analytical expansions, we shall see that, the discrepancies between two approaches (Gauss-Bonnet approach and the geodesic method) arises.

To analytically expand the hyper-elliptic integral expression of gravitational deflection angle in equation (\ref{gravitational deflection angle --- hyper-elipitical integral}), we assume that 
\begin{equation}
	\frac{1}{x_{0}} = \frac{2M}{r_{0}} \sim \frac{2M}{b} \ll 1
\end{equation}
It corresponds to the weak gravitational lensing where $M/b$ is not large. In theses cases, the gravitational deflection angle for acoustic Schwarzschild black hole can be expanded as follows
\begin{eqnarray}
	\alpha & = & \int_{0}^{1}
	\frac{dz}{\sqrt{P_{0}(x_{0})-P(z)}} -\pi \nonumber
	\\
	& = & 2\int_{0}^{1}
	\frac{dz}
	{\sqrt{\big[1-\frac{1+\xi}{x_{0}}+\frac{2\xi}{x_{0}^{2}}-\frac{\xi}{x_{0}^{3}}\big]-z^{2}\cdot\big[1-\frac{1+\xi}{x_{0}}\cdot z +\frac{2\xi}{x_{0}^{2}}\cdot z^{2} -\frac{\xi}{x_{0}^{3}}\cdot z^{3}\big]}} 
	-\pi \nonumber
	\\
	& = & 2\int_{0}^{1}
	\frac{dz}{\sqrt{1-z^{2}}} \cdot
	\frac{1}
	{\sqrt{1-\frac{1+\xi}{x_{0}}\cdot\frac{1-z^{3}}{1-z^{2}}+\frac{2\xi}{x_{0}^{2}}\cdot\frac{1-z^{4}}{1-z^{2}}-\frac{\xi}{x_{0}^{3}}\cdot\frac{1-z^{5}}{1-z^{2}}}}
	-\pi \nonumber
	\\
	& = & 2\int_{0}^{1}
	\frac{dz}{\sqrt{1-z^{2}}} \cdot 
	\bigg\{
	1+\frac{1+\xi}{2}\cdot\frac{1-z^{3}}{1-z^{2}} \cdot\frac{1}{x_{0}} 
	+\bigg[ \frac{3(1+\xi)^{2}}{8}\cdot\frac{(1-z^{3})^{2}}{(1-z^{2})^{2}}-\xi\cdot\frac{1-z^{4}}{1-z^{2}} \bigg] \cdot \frac{1}{x_{0}^{2}} \nonumber
	\\
	&   & \ \ \ \ \ \ \ 
	+\bigg[ \frac{5(1+\xi)^{3}}{16}\cdot\frac{(1-z^{3})^{3}}{(1-z^{2})^{3}} -\frac{3\xi(1+\xi)}{2}\cdot\frac{1-z^{3}}{1-z^{2}}\cdot\frac{1-z^{4}}{1-z^{2}}   
	+\frac{\xi}{2}\cdot\frac{1-z^{5}}{1-z^{2}} \bigg] \cdot \frac{1}{x_{0}^{3}}
	+O\bigg( \frac{1}{x_{0}} \bigg)^{4}
	\bigg\} 
	\ -\pi 
\end{eqnarray}
Finally, when $O$ and $S$ are both located at infinity, the gravitational deflection angle for acoustic Schwarzschild black hole can be expanded as
\begin{eqnarray}
	\alpha & = & \frac{2(1+\xi)}{x_{0}}
	+ \bigg[ \bigg( \frac{15\pi}{16}-1 \bigg) \cdot (1+\xi)^{2} - \frac{3\pi}{2} \cdot \xi \bigg] \cdot \frac{1}{x_{0}^{2}} \nonumber
	\\
	&   &  + \bigg[ 
	(1+\xi)^{3}\cdot\bigg(\frac{61}{12}-\frac{15\pi}{16}\bigg) + \xi(1+\xi)\cdot\bigg(\frac{3\pi}{2}-14\bigg) + \xi\cdot\frac{8}{3} \bigg] \cdot \frac{1}{x_{0}^{3}} 
	+ O\bigg( \frac{1}{x_{0}} \bigg)^{4} \label{gravitation deflection angle --- analytical expansion0}
\end{eqnarray}
where the following integrals have been used in the derivation
\begin{subequations}
	\begin{eqnarray}
		\int_{0}^{1}\frac{1}{\sqrt{1-z^{2}}}dz = \frac{\pi}{2} 
		\ \ & & \ \ 
		\int_{0}^{1}\frac{1}{\sqrt{1-z^{2}}}\frac{1-z^{3}}{1-z^{2}}dz = 2 
		\\
		\int_{0}^{1}\frac{1}{\sqrt{1-z^{2}}}\frac{1-z^{4}}{1-z^{2}}dz 
		= \frac{3\pi}{4}
		\ \ & & \ \ 
		\int_{0}^{1}\frac{1}{\sqrt{1-z^{2}}}\frac{1-z^{5}}{1-z^{2}}dz 
		= \frac{8}{3} 
		\\
		\int_{0}^{1}\frac{1}{\sqrt{1-z^{2}}} \frac{(1-z^{3})^{2}}{(1-z^{2})^{2}}dz 
		= \frac{5\pi}{4}-\frac{4}{3} 
		\ \ & & \ \ 
		\int_{0}^{1}\frac{1}{\sqrt{1-z^{2}}} \frac{(1-z^{3})^{3}}{(1-z^{2})^{3}}dz
		= -\frac{3\pi}{2}+\frac{122}{15}
		\\ 
		\int_{0}^{1}\frac{1}{\sqrt{1-z^{2}}} \frac{(1-z^{3})(1-z^{4})}{(1-z^{2})^{2}}dz
		= \frac{14}{3}-\frac{\pi}{2}
		\ \ & & \ \ 
	\end{eqnarray}
\end{subequations}
When the tuning parameter $\xi=0$, the result in equation (\ref{gravitation deflection angle --- analytical expansion0}) successfully reduces to the gravitational deflection angle for conventional Schwarzschild spacetime \cite{Virbhadra2000}
\begin{eqnarray}
	\alpha_{\text{Schwarzschild}} 
	& = & \frac{2}{x_{0}}
	+ \bigg( \frac{15\pi}{16}-1 \bigg) \cdot \frac{1}{x_{0}^{2}}
	+ \bigg( \frac{61}{12}-\frac{15\pi}{16} \bigg) \cdot \frac{1}{x_{0}^{3}}
	+ O\bigg( \frac{1}{x_{0}^{4}} \bigg) \nonumber
	\\
	& = & \frac{4M}{r_{0}}
	+ \bigg( \frac{15\pi}{16}-1 \bigg) \cdot \frac{4M^{2}}{r_{0}^{2}}
	+ \bigg( \frac{61}{12}-\frac{15\pi}{16} \bigg) \cdot \frac{8M^{3}}{r_{0}^{3}} 
	+ O\bigg( \frac{1}{r_{0}^{4}} \bigg)
\end{eqnarray}

However, in the above expansion in equation (\ref{gravitation deflection angle --- analytical expansion0}), the variable $x_{0}=r_{0}/2M$ is unknown. We should solve the minimal distance $r_{0}$ using the conserved quantities $J$, $E$ and impact parameter $b$. Following the notations in reference \cite{Weinberg1972}, the impact parameter is calculated through 
\begin{subequations}
\begin{eqnarray}
	& & b \equiv \frac{J}{\epsilon} = \frac{J}{1-E} 
	= \frac{r_{0}}{\sqrt{B(r_{0})}} = \frac{r_{0}}{\sqrt{f(r_{0})}} 
	= \frac{r_{0}}
	{\sqrt{1-\frac{2M(1+\xi)}{r_{0}}+\frac{8\xi M^{2}}{r_{0}^{2}}-\frac{8\xi M^{3}}{r_{0}^{3}}}}
	\\ 
	& \Rightarrow &
	\frac{2M}{b} = \frac{\sqrt{f(r_{0})}}{x_{0}}
	= \frac{\sqrt{1-\frac{1+\xi}{x_{0}}+\frac{2\xi}{x_{0}^{2}}-\frac{\xi}{x_{0}^3}}}{x_{0}} \label{equation impact parameter}
\end{eqnarray}
\end{subequations}
where we have used equation (\ref{conserved angular momentum}) and $E=0$ for particles traveling along null geodesics. We can expand both sides in equation (\ref{equation impact parameter}) into power series of $1/x_{0}$
\begin{subequations}
	\begin{eqnarray}
		\frac{2M}{b} & = & \frac{\sqrt{1-\frac{1+\xi}{x_{0}}+\frac{2\xi}{x_{0}^{2}}-\frac{\xi}{x_{0}^3}}}{x_{0}}  
		\approx 
		 \frac{1}{x_{0}} -\frac{1+\xi}{2}\cdot\frac{1}{x_{0}^{2}}
		-\frac{\xi^2-6\xi+1}{8}\cdot\frac{1}{x_{0}^3} 
		-\frac{\xi^{3}-5\xi^{2}+3\xi+1}{16} \cdot\frac{1}{x_{0}^{4}}
		+O\bigg( \frac{1}{x_{0}^{5}} \bigg)
		\\
		\bigg( \frac{2M}{b} \bigg)^{2} & = & \frac{1-\frac{1+\xi}{x_{0}}+\frac{2\xi}{x_{0}^{2}}-\frac{\xi}{x_{0}^3}}{x_{0}^{2}} 
		= \frac{1}{x_{0}^{2}} -(1+\xi)\cdot\frac{1}{x_{0}^{3}}
		+2\xi\cdot\frac{1}{x_{0}^{4}} -\xi\cdot\frac{1}{x_{0}^{5}}
		\\
		\bigg( \frac{2M}{b} \bigg)^{3} & = & \frac{\sqrt{\big(1-\frac{1+\xi}{x_{0}}+\frac{2\xi}{x_{0}^{2}}-\frac{\xi}{x_{0}^3}\big)^{3}}}{x_{0}^{3}} 
		\approx  
		 \frac{1}{x_{0}^{3}}
		-\frac{3(1+\xi)}{2}\cdot\frac{1}{x_{0}^{4}}
		+\frac{3\xi^{2}+10\xi+3}{8}\cdot\frac{1}{x_{0}^{5}}
		+O\bigg( \frac{1}{x_{0}^{6}} \bigg)
	\end{eqnarray}
\end{subequations}
Canceling the higher order terms of $1/x_{0}$, we finally obtain
\begin{equation}
	\frac{1}{x_{0}} 
	 =  \frac{2M}{b} 
	+\frac{1+\xi}{2} \cdot 
	\bigg( \frac{2M}{b} \bigg)^{2}
	+\frac{5\xi^{2}+2\xi+5}{8} \cdot 
	\bigg( \frac{2M}{b} \bigg)^{3} 
	+\frac{2\xi^{3}+\xi+2}{2} \cdot 
	\bigg( \frac{2M}{b} \bigg)^{4}
	+O\bigg( \frac{2M}{b} \bigg)^{5}
\end{equation}
In the expression of gravitational deflection angle (the equation (\ref{gravitation deflection angle --- analytical expansion0})), if we replace the $1/x_{0}$ by power expansions of $2M/b$, the deflection angle can be expressed into power series of $M/b$.
\begin{eqnarray}
	\alpha & = & \frac{2(1+\xi)}{x_{0}}
	+ \bigg[ \bigg( \frac{15\pi}{16}-1 \bigg) \cdot (1+\xi)^{2} - \frac{3\pi}{2} \cdot \xi \bigg] \cdot \frac{1}{x_{0}^{2}} \nonumber
	\\
	&   & + \bigg[ 
	(1+\xi)^{3}\cdot\bigg(\frac{61}{12}-\frac{15\pi}{16}\bigg) + \xi(1+\xi)\cdot\bigg(\frac{3\pi}{2}-14\bigg) + \xi\cdot\frac{8}{3} \bigg] \cdot \frac{1}{x_{0}^{3}} 
	+ O\bigg( \frac{1}{x_{0}} \bigg)^{4} \nonumber
	\\
	& = & \frac{4M(1+\xi)}{b}
	+\frac{3\pi M^{2}(5\xi^{2}+2\xi+5)}{4b^{2}}
	+\frac{8M^{3}(16\xi^{3}+8\xi+16)}{3b^{3}} 
	+O\bigg(\frac{M^{4}}{b^{4}}\bigg) 
	\label{deflection angle --- power series}
\end{eqnarray}
\end{widetext}
This result is slightly different from the power series obtained using Gauss-Bonnet theorem in equation (\ref{deflection angle vacuum}).

From the above power series of $M/b$, we can draw the conclusions. The power series of gravitational deflection angles obtained from the two approaches (the Gauss-Bonnet approach and the geodesic method) are consistent with each other only in lower-order contributions of $M/b$. In higher-order contributions of $M/b$, there are differences between the power series in equation (\ref{deflection angle --- power series}) and equation (\ref{deflection angle vacuum}). This is most probability caused by the approximations $\phi_{\text{observer}} \approx \pi + \alpha \approx \pi$ and $r(\gamma)=r(\phi)=b/\sin\phi$ used in the integration of Gauss curvature in subsection \ref{sec:4a}.

Furthermore, one can also derived the gravitational deflection angle for finite distance using the above geodesic method and expand the corresponding hyper-elliptical integrals into power series in a similar way. The procedure is tedious, we will not do this in the present work.







\end{document}